\documentclass[twocolumn,prd,showpacs,10pt,superscriptaddress,nofootinbib]{revtex4}
\usepackage{graphics}
\usepackage[pdftex]{graphicx}
\usepackage{amsmath, amssymb}
\usepackage{color}
\usepackage[usenames,dvipsnames]{xcolor}

\def\be{\begin{equation}}
\def\ee{\end{equation}}
\def\bea{\begin{eqnarray}}
\def\eea{\end{eqnarray}}
\def\ba{\begin{array}}
\def\ea{\end{array}}
\newcommand{\lsim}{\,\raise 0.4ex\hbox{$<$}\kern -0.8em\lower 0.62ex\hbox{$\sim$}\,}
\newcommand{\gsim}{\,\raise 0.4ex\hbox{$>$}\kern -0.7em\lower 0.62ex\hbox{$\sim$}\,}

\def\de{\mathrm{DE}}

\definecolor{MyBlue}{rgb}{0,0,1} 

\def\be{\begin{equation}}
\def\ee{\end{equation}}

\def\de F{\delta{F}}

\def\H0{H_{0}}

\def\de{\delta}

\definecolor{orange}{rgb}{1,0.5,0}

\begin{document}

\title{$f(R)$ gravity from the renormalisation group}
\date{March 16, 2012}

\newcommand{\addressSussex}{Department of Physics and Astronomy, University of Sussex, Brighton, BN1 9QH, United Kingdom}

\author{Mark Hindmarsh} 
\email{m.b.hindmarsh@sussex.ac.uk}
\affiliation{\addressSussex}
\affiliation{Helsinki Institute of Physics, P.O. Box 64, 00014 University of Helsinki, FInland}

\author{Ippocratis D. Saltas} 
\email{i.saltas@sussex.ac.uk}
\affiliation{Astronomy Centre, University of Sussex, Falmer, Brighton BN1 9QH, UK}

\begin{abstract}
We explore the cosmological dynamics of an effective $f(R)$ model 
constructed from a renormalisation group (RG) improvement of the Einstein--Hilbert action, using the non-perturbative beta functions of the exact renormalisation group equation. 
The resulting $f(R)$ model has some remarkable properties. It naturally exhibits an unstable de Sitter era in the ultraviolet (UV), dynamically connected to a stable de Sitter era in the IR, via a period of radiation and matter domination, thereby describing a non-singular universe.
We find that the UV de Sitter point is one of an infinite set, which make the UV RG fixed point inaccessible to classical cosmological evolution. In the vicinity of the fixed point, the model behaves as $R^2$ gravity, while it correctly recovers General Relativity at solar system scales. In this simplified model, the fluctuations are too large to be  the observed ones, and more ingredients in the action are needed. 
\end{abstract}


\maketitle


\section{Introduction}
Einstein gravity, although successful at solar and galactic scales, is challenged by cosmological observations of the early and late time Universe. Both early  (inflation)  and late time acceleration of our Universe require either the introduction of an extra degree of freedom in the action, like a scalar or tensor field, or a modification of gravity itself.
In particular, modifications of the gravitational action include scalar-tensor actions or non--linear extensions of Einstein--Hilbert action, probably the most famous being the so called Brans--Dicke \cite{Brans-Dicke} and $f(R)$ theories respectively \cite{StarobinskyInflation1, Nojiri&Odintsov_Review,Capozziello&Francaviglia_f(R)_Review,f(R)SotiriouFaraoni,f(R)Felice&Tsujikawa,MGSkordis}. A common characteristic among scalar-tensor and modified theories alike is that they lead to a modification of Newton's constant $G_{N}$, which acquires a scale dependence, for example in scalar-tensor theories through the coupling of gravity with a scalar field. 

At the same time, GR with a cosmological constant $\Lambda$ has been very successful in describing the late time acceleration of the Universe from a phenomenological point of view, but it is unable to account for a primordial inflationary era. Probably, the most challenging problems a cosmological constant faces from a theoretical point of view is the order of magnitude problem, i.e why it has such a tiny value, as well as the coincidence problem, or in other words why it is only at recent times that $\Lambda$ becomes dynamically relevant. In the context of scalar field or modified gravity models, the vacuum energy is replaced by a dynamically evolving, effective energy-momentum tensor, but this only partly solves the problem, as any effective energy-momentum tensor has to reproduce the tiny value of $\Lambda$ today.

In the last years, another modification of Einstein gravity has been suggested in the context of the Asymptotic Safety scenario \cite{Weinberg_AS}. According to this, Einstein gravity is quantisable as a metric theory in a non-perturbative way, provided a non-trivial fixed point exists at high energies, under the renormalisation group (RG) \cite{Litim_RG-Review, Nieder_RG-Review, Nieder-Reuter_RG_Review2, Percacci_RG-Review,Reuter-Sauer_RG-Review}. The existence of an ultraviolet (UV) fixed point ensures a well-defined UV behavior in the standard quantisation of gravity. Different investigations have shown that a non-trivial UV fixed point exists for the Einstein-Hilbert and higher truncations \cite{Narain-Perc_RG-ST1, Narain-Rahme_RG-ST2, Cod-Perc-Rahme1,Cod-Perc-Rahme2,Lit_FP}, 
providing strong evidence that quantum gravity is 
renormalisable in a non-perturbative way.

The starting point for this particular implementation of quantum gravity, is the scale dependent effective average action $\Gamma_{k}[g_{\mu \nu}]$, where $k$ is the cut-off energy, or coarse graining scale, which defines the scale 
above which all modes are integrated out, i.e.\ the functional integral is performed on
modes whose momentum satisfies $p^2 \ll k^2$. The effective action satisfies a functional renormalisation group equation \cite{Wett_RG-Eq} (see also \cite{Rosten:2010vm}), which describes the evolution of $\Gamma_{k}[g_{\mu \nu}]$ as a function of the scale $k$, giving rise to an RG flow on the space of different effective actions.  

Although any curvature invariant can be in principle included in the effective action, the actual study of the RG equations requires that a particular truncation is considered. In this paper, we will be focusing in the so-called Einstein-Hilbert truncation, where the only couplings are Newton's $G \equiv G_{k}(k)$ and the cosmological constant $\Lambda \equiv \Lambda_{k}(k)$, with $k$ the renormalisation group cut-off scale. The running of $G$ and $\Lambda$ changes the cosmological dynamics resulting from the action, and has also been suggested as a possible resolution to the coincidence problem \cite{Bon_Reut_RG-Cosmo, Bon_Reut_RG-Cosmo2, Wein, Tye_Xu, Bon_Cont, Cont, Reuter_RG-Astrophys, Grande_RGCosmo}. In particular, Refs \cite{Shap_Sola, Shap_Sola2,Shap_Sola3,Shap_Sola4} have studied the cosmological consequences of a running Newton's $G$, while Refs \cite{Shap_Sola_Stef, Bau, Bau2} studied the case of a running cosmological constant. 
Comparison
with cosmological observations, including supernovae data, has been carried out in Refs \cite{Gub_Horv_Stef, Shap_Sola5, Bonet_Lapuen_Shap_Sola, Shap_Sola_Stef}.

In a cosmological context,
it is attractive to think of the cut-off energy scale $k$ as dynamically evolving with cosmic time \cite{Bon_Reut_RG-Cosmo, Bonanno:2001hi, Reuter-Sauer_RG-Cosmo}.
There are different ways to understand this connection in an expanding Universe. Since in the effective action modes with momentum 
$p^2 \gg k^2$ are integrated out, 
$k$ defines the energy scale of the theory, i.e. the typical scale at which the couplings in the effective action are evaluated.

The typical energy of particles in an expanding Universe with temperature $T$ at a particular time, is directly linked with the expansion; the Universe starts off from a hot state and cools down as it expands with cosmic time $t$ and in particular, for a homogeneous and isotropic Universe described by the Friedman-Robertson-Walker (FRW) metric, characterised by the scale factor $a(t)$, the typical energy of relativistic particles scales as $1/a(t)$.
One could then think of identifying the cut-off scale $k$ as
\be
k \sim k_{B}T(t) \sim \frac{E_{0}}{a(t)}, \label{Ident-Temp}
\ee
where $k_{B}$ is the Boltzmann constant and $E_{0}$ a constant with dimensions of energy.

Alternatively, one can think that the horizon size of the Universe $d_{H} \sim 1/H(t)$, with $H\equiv \dot{a}(t)/a(t)$ the Hubble parameter, defines the typical scale of correlations between different quantum degrees of freedom, and identify
\be
k^{-1} \sim d_{H}(t) \sim H^{-1}(t). \label{Ident-H}
\ee
Notice that both identifications (\ref{Ident-Temp}) and (\ref{Ident-H}) are monotonically decreasing functions of cosmic time, in the context of a Hot Big Bang scenario. In addition, both of above identifications are performed at the level of the equations of motion, for which 
$k(t)$ is constrained through the Bianchi identities \cite{Bon_Reut_RG-Cosmo, Babic:2004ev, Reuter-Sauer_RG-Cosmo, Hind-Litim_Rahme, Reuter-Weyer_BD}, which provide the condition for all consistent identifications.

In this paper, we will consider a different identification which is performed at the level of the action, motivated by an analogous procedure which generates the effective potential for a scalar field theory. We will associate $k$ with the scalar curvature, i.e 
\be
k^2 \sim R,
\ee
through which we can view the RG improved Einstein-Hilbert action as an effective $f(R)$ model.  
The idea of Newton's $G$ running with curvature has been suggested previously \cite{Frol-RG-f(R)}, although not with the correct beta-function, and here we include the cosmological constant with the full non-perturbative beta functions for both couplings. The resulting $f(R)$ model does not include the renormalisation effects of matter, or any gravitational invariants other than $R$, and so it should be viewed as a prototype.  However,  it will turn out that it has some remarkable properties, also allowing us to study the RG improved action in an elegant way.  One feature is that the scale identification is performed at the level of the action, in a covariant fashion, so there is no need to add extra dynamical conditions through the Bianchi identities as described above. 

In the following, we will be working in a unit system with $c = \hbar = 1$, unless otherwise stated, as well as use $G = m_{p}^{-2}$, {$= 8\pi\kappa^2$}. Unless otherwise stated, mass scales will be presented in Planck units. 
    

\section{RG improved Einstein--Hilbert action}\label{sec:f(R)RG}
Our starting point is the RG improved effective action in the Einstein--Hilbert truncation, 
\be
\Gamma_{k}[g,\psi] = \int d^{4}x\sqrt{-g} \left[ \left(\frac{R(g) - 2\Lambda_{k} }{16 \pi G_{k}} \right) + \mathcal{L}_{\rm matter}(\psi, g)  \right], \label{RG_action_1}
\ee
with $k$ is the renormalisation group cut-off scale, which sets the momentum scale above which modes are integrated out. The effective, ``coarse-grained" action functional, $\Gamma_{k}[g,\psi]$, interpolates between the true effective action in the infrared (IR, $k \rightarrow 0$) and the bare action defined in the UV at a cut--off scale 
{$k_\textrm{max}$.} 
The interpolation of the effective action as a function of scale is controlled by the exact renormalisation group equation (ERGE)  \cite{Wett_RG-Eq}.
If 
{$k_\textrm{max}$} can be taken to infinity the theory is renormalisable, signaled by a UV fixed point in the couplings of the theory.

The quantum corrections can be encoded in the evolution of the coupling constants as a function of energy,\footnote{For a cautionary note see \cite{Anber:2011ut}.} whose beta-functions can be extracted from the ERGE. The form of the latter depends on the choice of the cut-off function choice and the gauge. We will follow the conventions of \cite{Litim_Opt-CO1,Litim_Opt-CO2}, noting that different choices of cut-off function and gauge do not change the qualitative features of the beta functions.

In the standard approach, one defines the dimensionless Newton's and cosmological constant as 
\be
g(k) \equiv k^{2}G(k)/{24\pi}, \; \; \; \lambda(k) \equiv \Lambda/k^2,
\ee
and the running of the dimensionless couplings in $d=4$ is described through the set of first order, coupled differential equations \cite{Litim:2003vp,Litim_RG-Review},
\begin{eqnarray}
&\partial_{t}\lambda = \beta_{\lambda}(g, \lambda) \equiv  -2\lambda - 12g - \frac{24g\left(3g + \frac{1}{2}(1-3\lambda) \right)}{2g - \frac{1}{2}(1-2\lambda)^2}, \label{betafunc1}\\
&\partial_{t}g = \beta_{g}(g, \lambda) \equiv 2g + \frac{24g^2}{4g - (1-2\lambda)^2},  \label{betafunc2}
\end{eqnarray}
where $t \equiv \ln k$, and $\beta_{\lambda},$ $\beta_{g}$ the beta functions. In above equations the factor of $24 \pi$ is included to remove phase space factors.

There are two fixed points of the above RG flows, a free or Gaussian one, with $(g^{*}, \lambda^{*})_{\rm GFP} = (0, 0)$, and an interacting one which is attractive in the UV ($k\to\infty$), with $(g^{*},\lambda^{*})_{\rm UV} = (0.015625,0.25)$ (Refs \cite{Litim_RG-Review}-\cite{Reuter-Sauer_RG-Review} and references therein).  The existence of a UV fixed point points to consistent quantum behavior of the system at high energies, realising Weinberg's Asymptotic Safety scenario \cite{Weinberg_AS}.  
The Gaussian fixed point ($k \rightarrow 0$) describes a free theory. 

A phenomenologically viable solution (trajectory) of the system (\ref{betafunc1})-(\ref{betafunc2}) on the $g-\lambda$ plane is one that starts at high energies from the UV fixed point and then evolves towards smaller values of $g$ as $k$ is lowered, passes close to the GFP, until it turns to the right towards increasing values of $\lambda$. 
A trajectory passing sufficiently close to the GFP will subsequently have a long classical regime, i.e $G \simeq G_{0}$, $\Lambda \simeq \Lambda_{0}$, with ``$0$" here denoting the present value.  The classical regime covering many orders of magnitude in scales is required by terrestrial, solar and galactic tests, as well as consistency with cosmological evolution since Big Bang Nucleosynthesis.

\begin{center}
\begin{figure*}
 \includegraphics[scale=1]{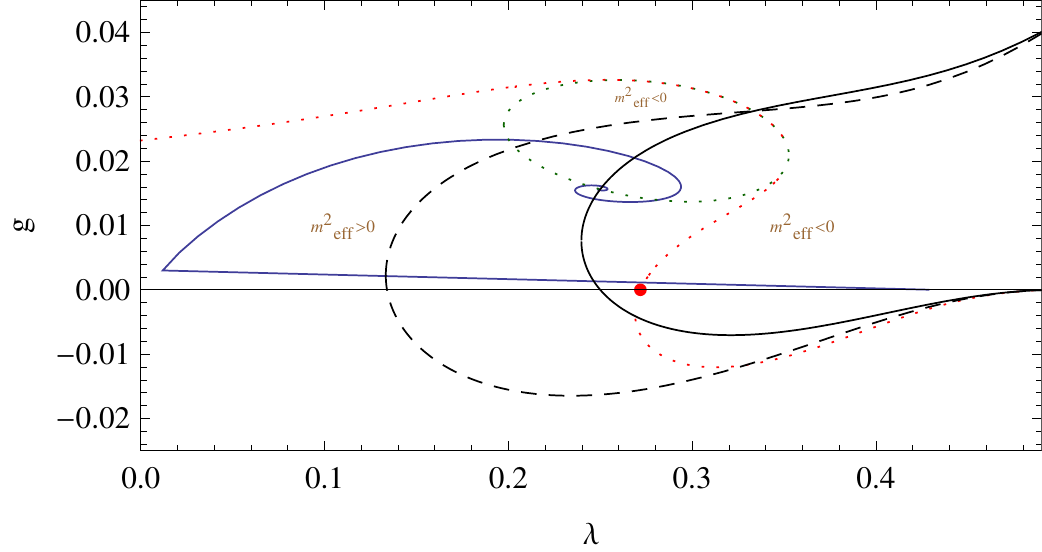} 
 \caption{\label{plot:PPhSpdSMassZero} A viable RG trajectory (blue) on the $g-\lambda$ plane in the Einstein--Hilbert truncation for the choice $\rho = 1$ and the initial conditions (\ref{cond:RG_IC}). It spirals around the UV RG fixed point and evolves towards the IR as curvature $R$ decreases. The intersection of the phase curve with the de Sitter line (black) corresponds to a {\it de Sitter point} in the cosmological evolution, while intersection with the dashed (black) one is where for the slow roll parameter $\epsilon_{V} = 1$. The regions where $m^{2}_{\rm eff} >0$ ($m^{2}_{\rm eff} <0$) are seperated by the dotted lines, with $m^{2}_{\rm eff}$ the Jordan frame mass squared, defined in (\ref{Scalar_Mass}). The dotted curve consists of two separate curves (green and red dots) corresponding to the vanishing of the numerator (denominator) of $m^{2}_{\rm eff}$. They join at the upper part of the dotted ``ellipsis", where  $m^{2}_{\rm eff}$ remains finite and non-zero. Along the lower part of the ``ellipsis" (green) $m^{2}_{\rm eff}$ vanishes. The dotted curves outside the ``ellipsis" (red) correspond to $m^{2}_{\rm eff} \rightarrow \infty$. Notice that beyond the red dot at $\lambda_{*} \simeq 0.27$ on the $\lambda$-axis, $m_{\rm eff}^2$ becomes negative, and therefore de Sitter space unstable too.} 
 \end{figure*}
 \end{center}

The Einstein-Hilbert truncation has a couple of features which may not be present in all truncations.
The eigenvalues of the linearised flow in the vicinity of the UV fixed point are complex conjugate, causing  oscillatory behavior of the trajectory around it (see Fig.\ \ref{plot:PPhSpdSMassZero}). 
Also, the flow (\ref{betafunc1},\ref{betafunc2}) has a singularity at $\lambda = 1/2$, which terminates the classical regime. It has been conjectured that this is an artifact of the truncation, and that there may actually be a non-trivial fixed point in the IR \cite{Bonanno:2001hi}.
\section{Scale identification \& effective $f(R)$ action}
As explained before, the first step in studying the cosmology of an RG improved action is to identify the cut-off scale $k$ with as a function of cosmic time $k=k(t)$. 

In this paper, we will work at the level of the action and use a particular ansatz that will allow us to view the effects of RG running of the couplings as an effective $f(R)$ model, by identifying 
\be
k^2 = \rho R, \label{Cut-Off-ident0}
\ee
where $R$ is the Ricci scalar and $\rho$ is a dimensionless constant. 
Here, with the particular identification (\ref{Cut-Off-ident0}) the dimensionless couplings are defined as $g(k) \equiv \rho R \times G(R), \; \; \; \lambda(R) \equiv \Lambda(R)/(\rho R)$, and action (\ref{RG_action_1}) takes the following form
\begin{align}
S_{f(R)} &=  \int d^{4}x \sqrt{-g} \frac{ R^2 h(R)}{384\pi^2}   + S_{\rm m}(\psi, g) \\
        &\equiv  \int d^{4}x  \sqrt{-g} \frac{f(R)}{2\tilde\kappa^2}  + S_{\rm m}(\psi, g), \label{RG_action_2}
\end{align}
with $h(R) \equiv \rho ( \frac{1- 2\rho \lambda}{g} )$, and the extra factor of $24 \pi$ appearing in the first line because of the rescaling of $g$ performed in the beta functions (\ref{betafunc1}) and (\ref{betafunc2}). 
We absorb it into the factor $\tilde\kappa^2 = 192\pi^2$.

The quantum corrections are now expressed in the 
non-linear effective action, which takes the form of an $f(R)$ model (\ref{RG_action_2}). This provides us with a different view of the RG effects on the Einstein--Hilbert action (\ref{RG_action_1}). What is more, the particular scale identification preserves general covariance of the action. 

{We can compare this procedure with the RG-improvement of the effective potential in scalar field theory \cite{Coleman:1973jx}.  There, if one starts with the tree potential $V= \lambda\phi^4/4!$, solves the RG equation for the coupling, and makes the identification $k = \alpha\phi$, one obtains at one loop
\be
V = \frac{1}{4!} \frac{\lambda_0}{1 - b(\lambda_0)\ln(\alpha\phi/k_0)}\phi^4,
\ee
where $b(\lambda) = {3\lambda}/{16\pi^2}$, which recovers and improves on the one-loop effective potential calculated by the standard graphical methods. The constants 
are constrained by a renormalisation condition such as 
\be
\left. \frac{d^4V}{d\phi^4}\right|_{\phi_0} = \lambda_\textrm{r},
\ee
where $\lambda_\textrm{r}$ is the physical coupling as inferred, say, from a scattering experiment with the background field set at $\phi_0$.
Normally, we can avoid all mention of $\alpha$ by writing $V=\lambda\phi^4/4!$, with  $\lambda=\lambda_\textrm{r}/(1 - b(\lambda_\textrm{r})\ln(\phi/\phi_0))$.  However, it is still implicit in the relationship between $\phi_0$ and the scale $k_0$.
}
 
The renormalisation conditions for the effective Einstein-Hilbert action can be taken as
\be
\label{e:RenPoi}
\left. f_R\right|_{R_0} = \frac{\tilde\kappa^2}{8\pi G_0}, \quad
\left. \frac{Rf_R -f}{2f_R}\right|_{R_0} = {\Lambda_0},
\ee
where $R_0$ is the curvature scalar evaluated today.

Some remarks regarding the action (\ref{RG_action_2}) are in order. Firstly, we can see that on a fixed point, where $h(R)$ is constant, the Lagragian is effectively $R^2$, which is renormalisable \cite{Stelle-R^2}.
Secondly, there is a singularity of the RG flow in the Einstein--Hilbert truncation \cite{Litim_RG-Review,Reuter-Sauer_RG-Cosmo}: the beta functions  
diverge for $4g = (\lambda -\frac{1}{2})^2$. 
We will therefore restrict ourselves to cosmological evolution which does not reach the singularity. 

Finally, let us comment on the dimensionless parameter $\rho$, 
defined through our identification (\ref{Cut-Off-ident0}), relating the RG scale $k$ and the cosmological scale $R$. We will see in Eq.\ (\ref{e:ScaMas}) that it determines the scalaron mass, and so in principle could also be fixed.  However, as we do not know the scalaron mass, we will leave $\rho$ free, and investigate what range of values give an acceptable cosmology.  As $\rho$ 
describes to what extent the RG scale $k$ follows the curvature $R$, we would hope to find that $\rho \sim 1$: it is natural to think of the RG scale as the scale of the important dynamics, which in the cosmological context is given by the curvature. It will in fact turn out $\rho \sim 1$ gives a viable cosmology.


\section{Stability and the GR limit} \label{sec:Stability}
%
\subsection{Degrees of freedom and stability conditions}
As a first step to understand the resulting effective $f(R)$ action from the renormalisation group (\ref{RG_action_2}), we want to study its stability and its approach to the limiting case of GR. Below we will introduce some basic facts about $f(R)$ gravity that will be necessary for the rest of the paper \cite{StarobinskyInflation1, Nojiri&Odintsov_Review,Capozziello&Francaviglia_f(R)_Review,f(R)SotiriouFaraoni,f(R)Felice&Tsujikawa,MGSkordis}.

It is well known that $f(R)$ models, exhibit an extra, massive scalar degree of freedom, dubbed ``scalaron". It satisfies a Klein--Gordon type equation, which can be found by varying action (\ref{RG_action_2}) with respect to the metric and then taking the trace,
\be
 \Box f_{R}(R) + \frac{dV_{\rm eff}(R)}{d f_{R}} = \frac{\tilde\kappa^2}{3} T_{\rm (m)}, \label{f(R)_Klein-Gordon}
\ee
where $\Box$ is the d'Alembertian associated to the metric $g_{\alpha \beta}$, $T_{\rm (m)} \equiv g^{\mu \nu}T_{\rm (m) \mu \nu}$ is the trace of any matter sources present, and 
\be
 \frac{d V_{\rm eff}(R)}{df_{R}}  \equiv \frac{1}{3}\left[ Rf_{R}(R) - 2f(R) \right]. \label{f(R)_Eff-Poten-Deriv}
 \ee
From (\ref{f(R)_Eff-Poten-Deriv}) we can deduce the scalaron's mass in the frame defined by action (\ref{RG_action_2}),
\be
m_{\rm eff}^2 \equiv \frac{d^{2}V_{\rm eff}(R)}{d f_{R}^{2}}  = \frac{ f_{R} - Rf_{RR} }{3 f_{RR}}. \label{Scalar_Mass}
\ee
Expression (\ref{Scalar_Mass}) also appears as the effective mass in a stability analysis around de Sitter spacetime \cite{f(R)SotiriouFaraoni} (and references therein).\footnote{Note that there is another definition for the scalaron mass in the Einstein frame, which we will present later.} 
Therefore, stability of the scalaron propagation (i.e avoidance of tachyonic instability), as well as stability of de Sitter spacetime requires that $m_{\rm eff}^2 > 0$.  

While an unstable scalaron just means that long-wavelength scalar fluctuations will grow, the graviton kinetic term must certainly have the correct sign, in order to avoid ghosts. This means that 
\be
f_R  > 0, \label{Condition_no-ghost_General}
\ee
which at small values of the couplings $\lambda$ and $g$, is ensured through condition 
\be
\frac{\rho R}{g} > 0. \label{Condition_no-ghost}
\ee
In order to make the connection with the RG, we will express both stability conditions, for the scalar and for the graviton, in terms of the beta functions, using the explicit form of action (\ref{RG_action_1}).

In the RG-improved Einstein--Hilbert action 
derivatives of $f$ can be expressed in terms of RG data, as 
\begin{align}
\frac{d}{dR}   
 = \frac{\partial }{\partial R} + \frac{1}{2R} \left(   \beta_{g}\frac{\partial }{\partial g} +  \beta_{\lambda} \frac{\partial }{\partial \lambda} \right),
\end{align}
For example, for $f_{R}$ we have, 
\be
f_{R} = 2R \left[ h - \frac{1}{4g}(h \beta_{g} + 2\rho^2 \beta_{\lambda}) \right] , \label{fR-RG}
\ee
while the second derivative is 
\begin{widetext}
\be
f_{RR} =  2h + \frac{\beta_{g}}{4g}\left( -6h + 2h\frac{\beta_{g}}{g} + 2\rho^2 \frac{\beta_{\lambda}}{g} - h\beta_{g,g} - 2 \rho^2 \beta_{\lambda,g} \right)  - \frac{\beta_{\lambda}}{4g}\left( 8+ 4 \rho^2 - 2\frac{\beta_{g}}{g} + h\beta_{g,\lambda} + 2 \rho^2 \beta_{\lambda,\lambda}\right). \label{fRR-RG}
\ee
\end{widetext}
Plugging above relations into expression (\ref{Scalar_Mass}), and using the beta functions (\ref{betafunc1},\ref{betafunc2}), the scalar mass $m_{\rm eff}^2$ can be re-expressed as $m_{\rm eff}^2 = m_{\rm eff}^2(R, g, \lambda)$.  The same is in principle true for $f_R$ and other quantities of interest as we will see later. 

In particular, from the explicit expression of $f_R$ in terms of the couplings, one can check that the no-ghost condition (\ref{Condition_no-ghost_General}) is always satisfied in the domain of interest, $0 < \lambda < 0.5$, $0 < g \lesssim 0.02$ and of course $R > 0$. (see also Fig.\ \ref{plot:PDfx2} for a plot of $f_R$ and Fig.\ \ref{plot:PPhSpdSMassZero} for the phase space of a viable RG trajectory.)

%
\subsection{The $f(R)$ model in the perturbative regime}
Let us now see how GR is recovered in this framework. If an RG trajectory is to be viable, it should have a sufficiently long classical regime, where any quantum corrections are suppressed enough not to be observed in astrophysical or solar system tests, and therefore the coupling constants should be effectively constant, and acquire the values observed at these scales. Therefore, GR is recovered in the sense that under the RG flow Newton's constant acquires its classical value, $G \simeq G_{0} = 1/m_{p}^2$ for a sufficient ``RG time", large enough to cover the range of classical scales (earth, solar and galactic). In the classical regime, $\Lambda$ has to have a negligible variation too. It has been shown that both requirements are achieved if the viable RG trajectory passes sufficiently close to the GFP at $(g, \lambda) = (0,0)$ \cite{Reuter_RG-Astrophys}. It is after the close passage to the GFP when the classical regime starts, and it turns out that the closer the trajectory passes to it, the longer it lasts in RG time, and the greater range of scales the classical regime covers. 

We will therefore need to first linearise the system of beta functions (\ref{betafunc1})-(\ref{betafunc2}) around the GFP \cite{Reuter_RG-Astrophys}. To make the analysis more clear, it would be better to first proceed with the linearisation of the equations without assuming any identification for $k$, i.e keeping $k$ as the independent variable in (\ref{betafunc1})-(\ref{betafunc2}). We get,
\begin{eqnarray}
&\partial_{t}\lambda = - 2\lambda + 2 \alpha g, \label{Linea_GFP1}\\
&\partial_{t}g  =  2g, \label{Linea_GFP2}
\end{eqnarray}
with renormalisation group time $t \equiv \ln (k/k_0)$, and $k_0$ a reference scale. The parameter $\alpha$ cut-off function dependent, but is always positive and of order 1. For the optimised cut--off, used to derive the beta functions (\ref{betafunc1})-(\ref{betafunc2}), $\alpha = 6$. 

The solution of the linearised system reads 
\begin{align}
& g = c_{1} k^2 \equiv g_{T} \frac{k^2}{k_{T}^2} \label{G_linear_k} \\
& \lambda = \frac{1}{2}\alpha c_1 k^2 + \frac{c_2}{k^2}  \equiv \frac{1}{2} \lambda_{T} \left( \frac{k^2}{k_T^2} + \frac{k_{T}^2}{k^2} \right), \label{Lambda_linear_k}
\end{align}
with $k_T$ the value of the cut-off scale around the turning point in the vicinity of the GFP, and $g_T \equiv g(k_T), \lambda_T \equiv g(k_T)$. Notice also that is,
\be
\lambda_T/g_T = \alpha, \label{lambda-gT}
\ee
which implies that $\lambda_T \sim g_T$, since $\alpha \sim O(1)$. From the above linearised relations we get for the dimensionful couplings,
\begin{align}
& \frac{8\pi G}{\tilde\kappa^2}= c_{1} = \frac{g_T}{k_T^2} = \mathrm{const.} \label{G_linear} \\
& \Lambda = \frac{1}{2}\alpha c_{1} k^4 + c_{2} = \frac{1}{2}\lambda_{T} \frac{k^4}{k_{T}^2} + \frac{1}{2}\lambda_{T}k_{T}^2, \label{Lambda_linear}
\end{align}
following the notation of Ref. \cite{Reuter_RG-Astrophys}.
Equation (\ref{G_linear}) tells us that in this regime, Newton's $G$ becomes a constant, and we identify $c_{1} = 8\pi G_{0}/\tilde\kappa^2$.
For scales $k \ll k_T$, $\Lambda$ is also effectively constant, and we may identify $c_2 = \Lambda_0$. 
Hence
\be
\frac{g_T \lambda_T}{2} =  \frac{8\pi G_0\Lambda_0}{\tilde\kappa^2},
\ee
and
\be
g_T = \sqrt\frac{16\pi G_0\Lambda_0}{\alpha\tilde\kappa^2},
\quad k_T = \left(\frac{\tilde\kappa^2\Lambda_0}{\alpha 4\pi  G_0}\right)^\frac14.
\ee
From the observed values of $\Lambda_0$ and $G_0$, we have 
\be
g_{T} \sim \lambda_{T} \sim 10^{-60}, \quad k_T \sim 10^{-30}m_\textrm{p}.
\ee
Let us now turn to the solution of the system under the identification $k^2 = \rho R$. 
The linearised equations are not enough as higher-order terms contribute already at O($R^2$).
An efficient way to include the higher-order terms is to substitute into the Talyor expansion around the renormalisation point $R_0$ (\ref{e:RenPoi})
\be
f(R) = f(R_0) + \left. f_R\right|_{R_0} (R-R_0) + \frac12 \left. f_{RR}\right|_{R_0} (R-R_0)^2.
\ee
We find that for the optimised cut-off where $\alpha = 6$, 
\be
f(R) \simeq  \frac{\tilde\kappa^2}{G_0}\left(R - 2 \Lambda_{0} \right) +  6 ( 2 - \rho) \rho  (R-R_0)^2. \label{f(R)_linear_correct}
\ee
From the small coupling expansion of (\ref{Scalar_Mass}) the scalaron mass squared in the classical regime is given by 
\be
m_{\rm eff,0}^2 = \left. \frac{f_R-Rf_{RR}}{3 f_{RR}}\right|_{R_0} \simeq 
 \frac{1}{36 (2 - \rho ) }\frac{R_0}{g},
\ee 
(see also Eq.\ \ref{relation:Scal_mass_gsmall}) and we see that it is positive provided $0 < \rho < 2$. 
Using the renormalisation condition (\ref{e:RenPoi}) we find 
\be
\label{e:ScaMas}
m_{\rm eff,0}^2 \simeq \frac{1}{36 (2 - \rho )}\frac{\tilde\kappa^2}{8\pi G_0},
\ee 
and observe that the scalaron mass is safely at the Planck scale, so large deviations from GR at laboratory, solar and astrophysical scales are avoided.

To get an idea of the realistic values of the couplings in the classical regime we can evaluate them at solar and galactic scales,  taking $k^2 \sim R$. With $R_{\rm sol}^{-1/2} \sim1 {\rm AU}$ 
\footnote{$1 {\rm AU} \simeq 1.496 \times 10^{11}$ m.} 
and  $R_{\rm gal}^{-1/2} \sim 10^{21} {\rm m}$ we find that
\be
g_{\rm sol} \simeq R_{\rm sol} \times G_{\rm sol} \simeq 10^{-92}, \; \; g_{\rm gal} \simeq R_{\rm gal} \times G_{\rm gal} \simeq 10^{-112}, \label{Solar-Galact-Values}
\ee
assuming that $G_{\rm gal}  = G_{\rm sol}  \simeq 10^{-70}\; {\rm m}^{2}$.
We see that the classical value of the dimensionless coupling $g$ acquires a tiny value. For $\lambda$ we cannot follow the same analysis, since $\Lambda$ has been only measured at cosmological scales, $k \sim H_{0}$, with $H_{0}$ the Hubble parameter today. 
However, the product $g\lambda \sim G_0\Lambda_0$, so $\lambda_{\rm sol} \ll 1$.
Therefore, the values of both $g$ and $\lambda$ on solar and galactic scales lie extremely close to the GFP.  However, it is intriguing to note that by this reasoning, $\lambda$ evaluated at the Hubble scale is of order 1, where non-perturbative effects in the beta functions are important \cite{Reuter_RG-Astrophys}.

\begin{center}
\begin{figure}
\includegraphics[scale=0.75]{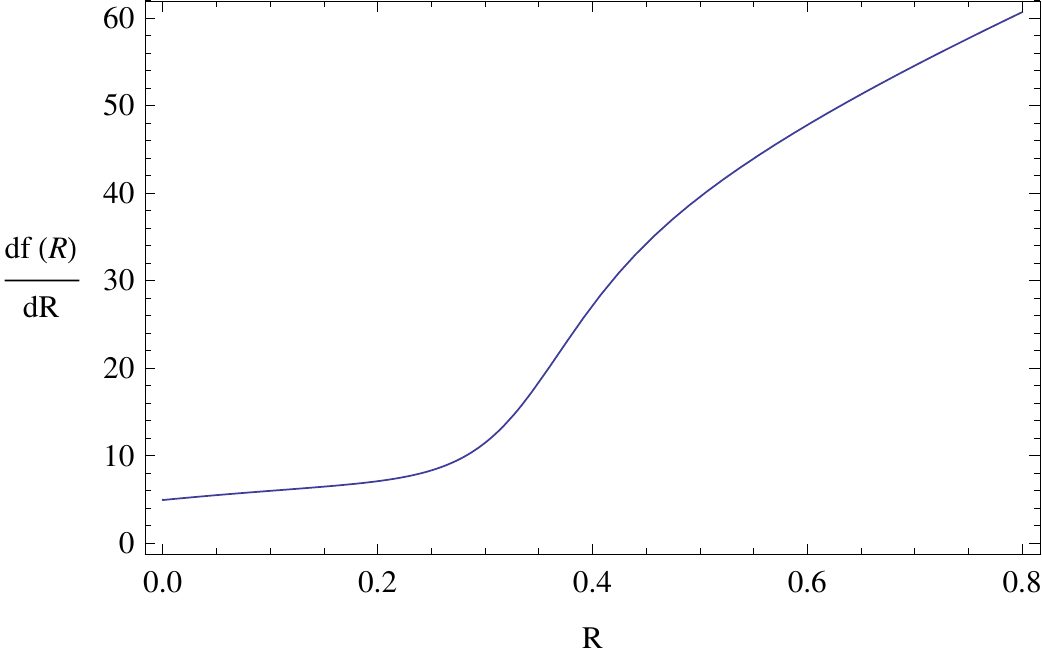} 
\caption{ \label{plot:PDfx2} The derivative of the $f(R)$ model for the initial conditions (\ref{cond:RG_IC}) (in Planck units). For large values of $R$, the model effectively behaves as $R^2$ gravity, since $f_{R}(R) \sim R$. For smaller $R$, the evolution enters the classical regime near the Gaussian fixed point. The Einstein-Hilbert term dominates, $f_{R}(R)$ becomes nearly constant, with a small positive slope reflecting the positivity of the scalaron mass-squared. At very small $R$ (not shown) there is a departure from the Einstein-Hilbert action due to the IR divergence in the beta functions. This part of the action is never encountered as the system freezes at the IR de Sitter point.} 
\end{figure}
\end{center}

The form of a phenomenologically viable RG evolution on the $g-\lambda$ plane is given in Fig.\ \ref{plot:PPhSpdSMassZero}, for the choice of $\rho = 1$ and 
\footnote{Numerical solutions in this paper are obtained using Mathematica's differential and algebraic solvers, making use of the stiffness option as well as increasing the maximum step number when appropriate. Plots are also produced with the same software.}
\begin{align}
&R_{\rm min} = 8\times10^{-5}, \; R_{T} = 5\times10^{-3}, \;R_{\rm max} = 50, \nonumber \\
&\lambda(R_{T})= 10^{-2}, \; g(R_{T})= 10^{-3}, \; \label{cond:RG_IC}
\end{align}
where $R_{T}$ is the curvature at the turning point close to the GFP, in Planck units. The above initial conditions are not realistic, but they allow for a good numerical illustration. Fig.\  \ref{plot:PDfx2} shows the evolution of the derivative $f_{R}(R)$ of the resulting $f(R)$ model, for the above choice of initial conditions. 

\section{Cosmological dynamics}

Now, we proceed with studying the cosmological dynamics of the model, i.e the cosmological fixed points, their stability and the transition from one cosmological era to the other, as well as if inflation can be viable in this scenario. 

A viable cosmological model, aiming to describe the background evolution of the universe from early to late times, should have a period of accelerated expansion at early times (inflation), followed by a radiation and matter era respectively, evolving asymptotically towards de Sitter at late times. Each particular period has its own requirements in order to be viable. For example, a UV de Sitter point should be unstable, while an IR one stable, while the matter point should be a saddle with damped oscillation, so that strucutre formation has enough time to take place. 

\subsection{Transition to the Einstein frame}

It will be useful for the latter analysis to first calculate the Einstein frame action, as an aid in calculating inflationary quantities, like the slow roll parameters. To do this we will  introduce an auxiliary field and then conformally transform the metric appropriately. However, in the context of action (\ref{RG_action_2}) the latter transformation requires care, since Newton's $G$ is running with curvature.

Let us see this in more detail. Our original action (\ref{RG_action_2}) is a function of $R, g(R), \lambda(R)$ and implicitly of the metric through the Ricci scalar $R$. In the standard way,  we introduce auxiliary scalars $\sigma$ and $\phi$, and write our original theory in the Jordan frame as
\begin{align}
&S = \frac{1}{2\tilde\kappa^2}\int \sqrt{-g} \left[ f'(\sigma) R - (f'(\sigma) \sigma - f(\sigma)) \right] \\
& \equiv  \frac{1}{2\tilde\kappa^2}\int \sqrt{-g} \left[ \phi R - V(\phi) \right], \label{RG_action_2_Jordan}
\end{align}
with $V(\phi) = \phi \sigma(\phi) - f( \sigma(\phi) )$, and $\phi = f'(\sigma)$. We require that $f''(\sigma) \neq 0$, so that the function $f'$ can be inverted to find $\sigma$ as a function of $\phi$.
Note that the equation of motion for $\sigma$ gives the constraint which reproduces the original action, i.e $\sigma = R$.

The Jordan frame scalar $\phi$, plays the role of the inverse of Newton's constant in front of $R$. For the transition to the Einstein frame, Newton's constant will have to be re-introduced through the conformal redefinition of the metric, and the question that arises in our scenario, is which Newton's constant should that be, since using a running $G = G(R)$ could lead to ambiguities. We can resolve this issue by using Newton's $G$ today, denoted $G = G_{0}$.

We can now perform the conformal redefinition of the metric as
\be
\tilde{g}_{\alpha \beta} = \frac{8\pi G_{0}\phi }{\tilde\kappa^2}g_{\alpha \beta},
\ee
combined with a redefinition on the scalar $\phi$
\be
\phi = \phi_{0} \exp \left(\sqrt{\frac{16 \pi G_{0}}{3}} \Phi \right),
\ee
with $\phi_{0}$ constant. 
Performing above two field redefinitions in action (\ref{RG_action_2_Jordan}), we finally end up with the Einstein frame action
\begin{widetext}
\be
\hspace{-0.5cm} \widetilde{S} = \int d^{4}x \sqrt{-\tilde{g}} \left( \frac{1}{16 \pi G_{0}}\widetilde{R} - \frac{1}{2} (\nabla \Phi)^2 - U(\Phi) \right) + \widetilde{S}_{m}(\tilde{g}, \psi, \Phi).
\ee
\end{widetext}

The scalaron potential in the Einstein frame is then given in parametric form as,
\begin{align}
&U(R) = \frac{\tilde\kappa^2}{2\left( 8\pi G_{0} \right)^2} \frac{R f_{R}(R) - f(R)}{ f_{R}(R)^2}, \label{Potential_Einstein} \\
& \Phi(R)  = \sqrt{ \frac{3}{16 \pi G_{0}} } \ln f_{R}(R). \label{Scalar_Einstein}
\end{align} 

\begin{center}
\begin{figure}
 \includegraphics[scale=0.75]{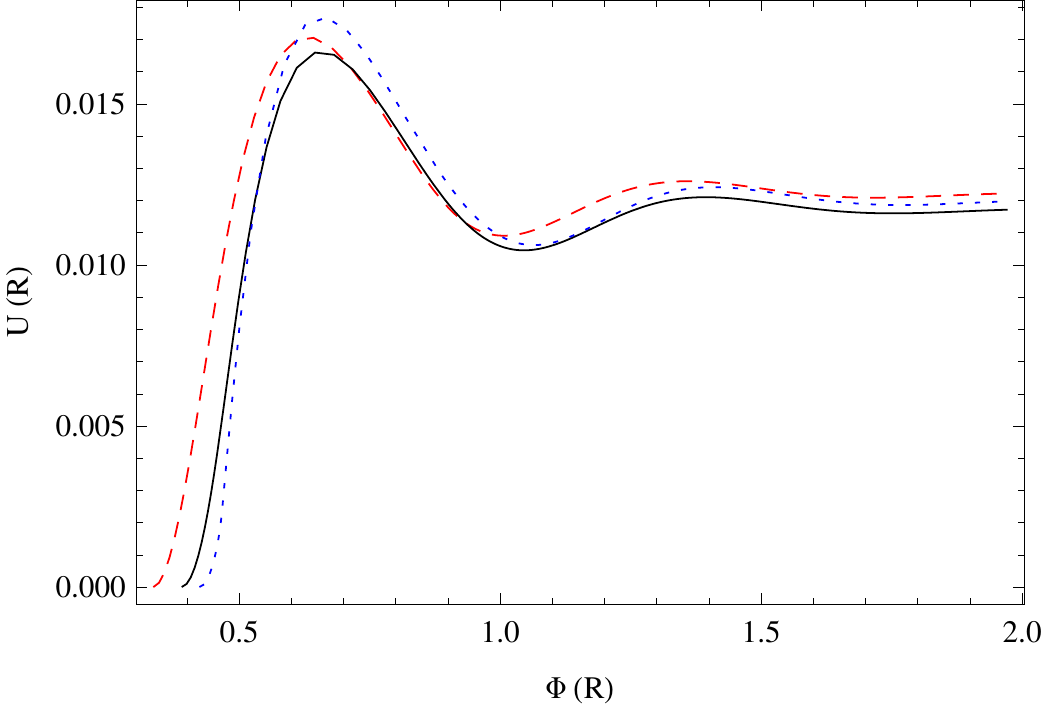} 
 \caption{ \label{plot:PPotentialEinsteinMulti}The Einstein frame scalar potential (in Planck units), described by relations (\ref{Potential_Einstein})-(\ref{Scalar_Einstein}), for $\rho = 0.8$ (red, dashed), $\rho=1$ (back, continuous) and $\rho=1.15$ (blue, dotted) respectively. Cosmological evolution starts from the maximum of the potential, which corresponds to the unstable UV de Sitter point, and evolves towards smaller values of the field $\Phi$.}
  \end{figure}
\end{center}

The mass of the scalar $\Phi$ is defined in the usual way through the Einstein frame potential as
\be
\widetilde{m}^{2}_{\rm eff} = \frac{d^{2}U}{d \Phi^2}.
\ee
We will use the above two relations later when we will work out the inflationary slow roll parameters. The Einstein frame potential for different values of $\rho$ is plotted in Fig.\ \ref{plot:PPotentialEinsteinMulti}. The maximum corresponds to the unstable de Sitter point in the UV, with the cosmological evolution occurring ``to the left" of it, i.e to smaller field values. 

\subsection{de Sitter solutions} \label{sec:deSitter}
Let us now look for the simplest cosmological solutions, which are the maximally symmetric constant curvature ones. 
In $f(R)$ gravity, they correspond to the points where the potential in (\ref{f(R)_Klein-Gordon}) has an extremum, i.e solution of the algebraic equation
\be
Rf_{R}(R) - 2f(R) = 0. \label{dS_Condition}
\ee
One can check that the same condition is also derived in the Einstein frame by requiring that $dU/d\Phi = 0$.

Using relation (\ref{fR-RG}), condition (\ref{dS_Condition}) implies that
\begin{equation}
2 \rho^2 g \beta_{\lambda} + (1-2 \rho \lambda)\beta_{g} = 0. \label{dSp_beta}
\end{equation}

For $\beta_{\lambda}$, $\beta_{g}$ non-zero, and a given $\rho$, equation (\ref{dSp_beta}) defines a family of solutions, described by a curve in the $g-\lambda$ plane, which is the locus of all de Sitter points. Any intersection of it with the RG trajectory will imply a de Sitter era in the particular cosmological evolution. It is interesting to note that any RG fixed point will always satisfy the de Sitter condition (\ref{dS_Condition}) or (\ref{dSp_beta}), since there, $\beta_{g} = \beta_{\lambda} =0$. This is an identity for fixed points, as $f(R) \propto R^2$ there, but we can check that they are de Sitter by inspecting the Einstein frame potential.  
In particular, the UV RG fixed point is a always a de Sitter point, as the potential (\ref{Potential_Einstein}) stays finite as $R\rightarrow \infty$. 

The location of the de Sitter line depends on the value of the parameter $\rho$, which shifts the scale of both early and late time de Sitter points. As a starting point, we can get an idea of the de Sitter points structure by setting $\rho = 1$ in equation (\ref{dSp_beta}) and working out the resulting de Sitter line, which is shown in Fig.\  \ref{plot:PPhSpdSMassZero}. The de Sitter line passes through the UV RG fixed point, yielding this way an infinite number of de Sitter points. This can be seen as follows:  as pointed out before, the RG UV fixed point is a de Sitter point itself. On the same time, the behavior of the RG evolution in the vicinity of the RG UV fixed point is described by an unstable spiral, which circles the fixed point infinitely many times as $k \rightarrow \infty$ (or $R \rightarrow \infty$). As a consequence there will be an infinite number of intersections between the de Sitter line and the RG phase curve. The UV RG fixed point is the limiting de Sitter point of the above infinite set of de Sitter points.

Furthermore, as Fig.\  \ref{plot:PPhSpdSMassZero} shows, there is an ``outer" de Sitter point in the UV regime, and another one in the IR. For the case $\rho = 1$, we find
\be
(g_{\rm dS}, \lambda_{\rm dS})_{\rm UV} \simeq (0.02, 0.27),
\ee
while it is easy to show that 
\be
(g_{\rm dS}, \lambda_{\rm dS})_{\rm IR} \simeq (0,0.25).
\ee
Notice that the ``inner'' UV de Sitter points cannot be accessed, since they are protected by the outer one. At least in the Einstein--Hilbert truncation, and under the cut--off identification considered here ($k^2 = \rho R$), this seems to be a general behavior: There is always an infinite set of UV de Sitter points, all hidden by the most outer one, and a de Sitter point in the IR. As a consequence, classical cosmological evolution cannot reach the extreme UV regime around the UV RG fixed point, i.e for $k^2 = \rho R \rightarrow \infty$. 

We now want to understand how the de Sitter line changes as we vary the dimensionless parameter $\rho$ in our cut--off identification. There are two extreme cases leading to two limiting de Sitter lines, one for $\rho \rightarrow \infty$ and another for $\rho \rightarrow 0$. Solving equation (\ref{dSp_beta}) for $g$ and taking the limit $\rho \rightarrow \infty$ the limiting curve is described by
\begin{widetext}
\begin{align}
&g_{\rho\rightarrow \infty}(\lambda) = 
\frac{1}{96} \left(12 \lambda ^2 -4 \lambda -3 + \sqrt{144 \lambda ^4+672 \lambda ^3-824 \lambda ^2+216 \lambda +9} \right), \label{dSLimitingCurveLargeRho}
\end{align}
\end{widetext}
which for a realistic RG evolution gives a de Sitter point at the UV, and another one very close to the GFP, i.e $\lambda \sim g \sim 10^{-35}$. Therefore, by tuning the parameter $\rho$ to very large values, both UV and IR de Sitter points are shifted towards the UV along the RG trajectories. Notice that letting $\lambda \to 0$ in (\ref{dSLimitingCurveLargeRho}) we get $g_{\rho\rightarrow \infty} \to 0$, i.e the curve passes through the GFP at $(\lambda, g) = (0, 0)$.

On the other hand, as $\rho \rightarrow 0$, the de Sitter line becomes
\[
g_{\rho\rightarrow 0}(\lambda) = \frac{1}{16} (1-2 \lambda )^2,
\]
which again gives a de Sitter point in the UV and a second one for $(g,\lambda) \simeq (0,0.5)$, as $g  \ll 1$ in the IR regime, but now both points are shifted towards the IR. The de Sitter lines corresponding to the extreme cases described above can be seen in Fig.\  \ref{plot:PLimdSEpsilonV}.

To summarise: the general trend is that by making $\rho$ smaller, the position of the UV de Sitter point is shifted towards smaller values of $R$ (i.e moving away from the RG UV fixed point), while the situation is the opposite for increasing $\rho$. 
\begin{center}
\begin{figure}
 \includegraphics[scale=0.65]{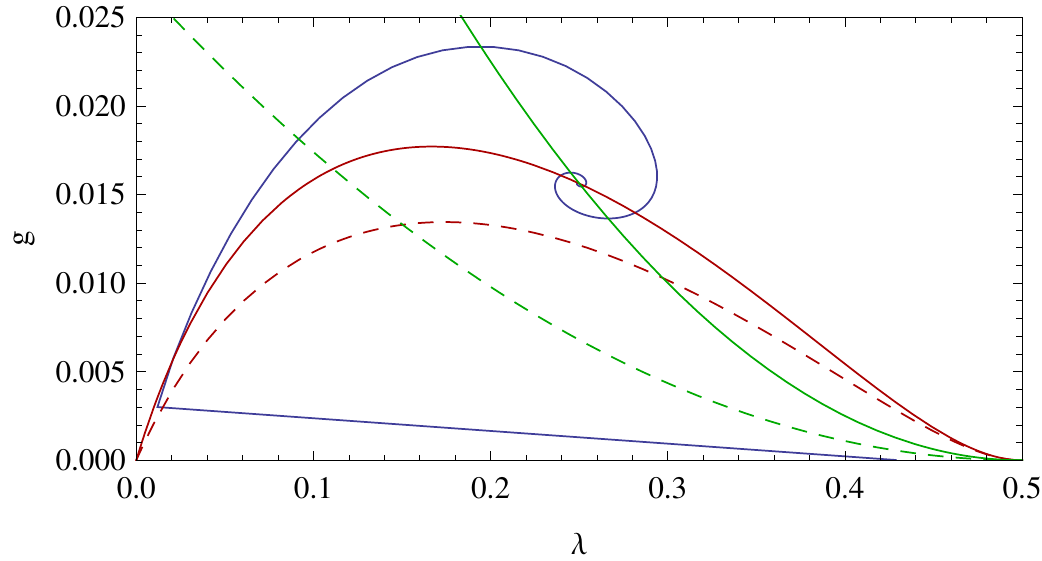}  \label{plot:PLimdSEpsilonV}
 \caption{\label{plot:PLimdSEpsilonV} The limiting de Sitter (continuous) and slow roll lines (dashed) respectively. Red colour corresponds to $\rho \rightarrow \infty$ (``bell" shaped curves), and the green to $\rho \rightarrow 0$ respectively. For $\rho \rightarrow \infty$ both de Sitter and slow roll line go to zero as $\lambda \rightarrow 0$.}
 \end{figure}
\end{center}
Let us turn attention to the stability of the de Sitter points. As said before, a de Sitter point is (un)stable if ($m_{\rm eff}^2<0$) $m_{\rm eff}^2 > 0$. Therefore, the equation 
\[
m_{\rm eff}^2(R, g, \lambda)= 0 
\]
will in turn define a line on the $g-\lambda$ plane along which the square of the mass becomes zero. Another useful line on the $g-\lambda$ plane is the one along which the square mass diverges, i.e its denominator becoming zero. Both $m_{\rm eff}^2 = 0$, and $m_{\rm eff}^2 \to \infty$ lines will divide the $g-\lambda$ plane into regions of positive and negative mass squared. This can be seen in Fig.\  \ref{plot:PPhSpdSMassZero} for $\rho =1$. If the UV de Sitter point should play the role of an inflationary era, it should be an unstable, while the IR one should be stable. We will see later that this can be achieved for a range of values for the parameter $\rho$.

The general expression for the scalaron mass in terms of the dimensionless couplings is quite complicated, but it simplifies reasonably in the classical and IR regime, where $g \ll 1$. In this case, the mass takes the form
\be
m_{\rm eff (IR)}^2 \simeq \frac{{R}(2 \lambda -1)^3}{36g\left( 8 \lambda ^3 \rho +4 \lambda ^2 \rho +\lambda  (8-6 \rho )+\rho -2\right )}, \label{relation:Scal_mass_gsmall}
\ee
where we neglected terms of order $g^2$ and higher. The critical points where the denominator of the scalaron mass vanishes will signal a singularity, with the scalaron mass going to infinity. 

Let us study the positivity of $m_{\rm eff (IR)}^2$ by first studying the special case of the GFP regime, where in addition to $g \ll 1$, it is also $\lambda \ll 1$. In this case, relation (\ref{relation:Scal_mass_gsmall}) simplifies to 
\be
m_{\rm eff (GFP)}^2 \simeq \frac{R}{36 g \left( 2 - \rho \right)}, \label{relation:Scal_mass_IR_GFP}
\ee
and the sign of it is positive when $\rho < 2$, while for $\rho \to 2$ it blows up. 
We recall that the renormalisation condition (\ref{e:RenPoi}) fixes $\rho R_0/g_0 \sim m_{\rm P}^2$, so the scalaron has a Planck-scale mass near the Gaussian fixed point.

In the IR regime, where $\lambda \sim O(1)$, we have to study the full relation (\ref{relation:Scal_mass_gsmall}). 
We distinguish two regimes, one when $\rho<2$ and another when $\rho> 2$. For $\rho < 2$, the vanishing of the denominator of (\ref{relation:Scal_mass_gsmall}) has only one relevant solution $\lambda = \lambda_{*}(\rho)$ being a function of $\rho$.
\begin{align}
& 0 < \lambda < \lambda_{*} : m^2_{\rm eff} > 0 ,\\
& \lambda_{*} < \lambda <  0.5 : m^2_{\rm eff} < 0.
\end{align}
In the limiting case of $\rho \rightarrow 0$, $\lambda_{*} \rightarrow 0$, while as $\rho \rightarrow 2^{-}$, it is $\lambda_{*} \simeq 0.31$. 

On the other hand, for $\rho > 2$, there are two relevant solutions, $\lambda_{*(1)}$ and $\lambda_{*(2)}$. We have the following cases
\begin{align}
&0 <\lambda< \lambda_{*(1)} : m^2_{\rm eff} < 0,\\
&  \lambda_{*(1)} < \lambda <  \lambda_{*(2)} : m^2_{\rm eff} > 0, \\
&\lambda_{*(2)} <\lambda < 0.5 : m^2_{\rm eff} < 0,
\end{align}
with both $\lambda_{*(1)}$, $\lambda_{*(2)}$ varying with $\rho$, i.e $\lambda_{*(1)} \equiv \lambda_{*(1)}(\rho)$, $\lambda_{*(2)}\equiv \lambda_{*(2)}(\rho)$. In particular, we have that as $\rho \rightarrow 2^{+}$ $\lambda_{*(1)} = 0$ and $\lambda_{*(2)} \simeq 0.31$, while for $\rho \rightarrow \infty$ $\lambda_{*(1)} \simeq 0.2$ and $\lambda_{*(2)} \simeq 0.5$.

We conclude from the above analysis that the case $\rho > 2$ is rejected, since $m_{\rm eff}^2$ is negative around the GFP. On the other hand, for $\rho < 2$ the mass $m_{\rm eff}^2$ is positive around the GFP ($\lambda \ll 1$) and it stays positive for $\lambda < \lambda_{*}$ with $\lambda_{*}$ approaching $\lambda_{*} \simeq 0.3$ as $\rho \rightarrow 2$.
 
An important point when $\rho <2$, concerns the position of the IR de Sitter point. From above, it is understood that both the position of de Sitter points as well as the critical point $\lambda_{*}(\rho)$, beyond which $m_{\rm eff}$ becomes negative, depend on the parameter $\rho$. What it turns out to be is that the corresponding position of the IR de Sitter point, will lie ahead of $\lambda_{*}$ on the $\lambda$-axis for $\rho \lesssim 0.9$, which means that the de Sitter point will be unstable. As a result, all trajectories with $g \ll 1$ and $\rho \lesssim 0.9$, will posses an unstable IR de Sitter point. In other words, the RG trajectory will pass through the mass singularity point $\lambda_{*}$, making $m^2_{\rm eff}$ negative, before the trajectory reaches its actual terminating (de Sitter) point. 

From the above stability analysis, we see that the parameter $\rho$ has been constrained to be $0.9 \lesssim \rho < 2$. In the next section, we will further constraint $\rho$ by requiring that the different cosmological periods are connected with each other in a viable way, finding that $\rho \sim 1$.


\subsection{Dynamical evolution from UV to IR}\label{dynamical-system}
We saw that in principle we can have de Sitter solutions, and the existence of a classical regime ensures for a standard radiation/matter era respectively. It is important though, that the cosmological eras are connected dynamically in a viable way. This will be the subject of this section. More precisely, we will consider action (\ref{RG_action_2}) in the presence of a perfect fluid with barotropic index $w \equiv p/\rho$, and study its dynamics by means of a dynamical system analysis, by improving the dynamical system for $f(R)$ gravity, presented in Ref. \cite{Amendola_fR-DS}, to account for our RG-inspired $f(R)$ model.

We can start by defining the following dimensionless variables,
\begin{align}
& x_{1} = \frac{- \dot{f_{R}}}{H f_{R}}, \label{x1}\\
& x_{2} = \frac{-f}{6 H^2f_{R}}, \label{x2}\\
& x_{3} = \frac{R}{6H^2}, \label{x3} \\
& x_{4} = \frac{\tilde{\kappa}^2 \rho_{r} }{ 3 H^2 f_{R} },
\end{align}
with an over dot denoting derivative with respect to cosmic time. The Hubble parameter is defined as $H \equiv \dot{a}/a$, with $a$ the Universe scale factor. In the absence of radiation it is $x_{4} = 0$.

Then, the background dynamics can be expressed in terms of the dynamical system \cite{Amendola_fR-DS},
\begin{align}
&x_{1}' = -1 - x_{2} - 3x_{2} + x_{1}^2 - x_{1}x_{3} + x_{4}, \label{deq1} \\
& x_{2}' = \frac{x_{1}}{x_{3}} - x_{2}(2x_{3}  -x_{1} - 4),\label{deq2} \\
& x_{3}' = \frac{-x_{1}x_{3}}{m} - 2x_{3}(x_{3} - 2),\label{deq3} \\
&x_{4}' = - 2x_{3}x_{4} + x_{1}x_{4},\label{deq4} 
\end{align}
with the constraint 
\be
\Omega_{m} \equiv \frac{\tilde{\kappa}^2 \rho_{m}}{3 H^2 f_{R}} = 1- x_{1}- x_{2}-x_{3} - x_{4},
\ee
and primes here denoting differentiation with respect to $\ln a$.

The quantity $m = m(r)$ is defined as 
\begin{align}
& m \equiv \frac{d \ln f_{R}}{d \ln R} = \frac{R f_{RR}}{f_{R}}, \label{m(r)} \\
& r \equiv - \frac{d \ln f}{d \ln R} = - \frac{R f_{R}}{f} = \frac{x_{3}}{x_{2}}. \label{r}
\end{align}
$m = m(r)$ characterizes the particular $f(R)$ model, and it needs to be given a priori in order for the dynamical system to close.
In principle, given a particular $f(R)$ model, one is able to invert $r = r(R)$ and plug into $m$ to get $m = m(r)$. However, in our case the form of the $f(R)$ model is dictated through the particular running of the couplings $g(R)$, $\lambda(R)$, by solving the system of beta functions. Therefore, in order to close the dynamical system (\ref{deq1})-(\ref{deq4}) we will need to evolve the couplings with time as well. 

In addition, the effective equation of state is given by,
\be
w_{\rm eff} = -\frac{1}{3}\left( 2x_{3} -1 \right). \label{weff}
\ee

For the dimensionless couplings we can write,
\begin{align}
& g ' = \frac{\partial g }{ \partial R}  \frac{dR}{d r} \frac{d r}{dN}  = \frac{ \beta_{g} }{2R} \frac{\partial R}{ \partial r} \left( \frac{\partial r}{ \partial x_{2} } x_{2}' +   \frac{\partial r}{\partial x_{3} } x_{3}'  \right), \\
& \lambda ' = \frac{\partial \lambda }{d R} \frac{d R}{ \partial r} \frac{d r}{dN}  = \frac{\beta_{\lambda} }{2R} \frac{\partial R}{ \partial r} \left( \frac{\partial r}{\partial x_{2}} x_{2}' +   \frac{\partial r}{\partial x_{3}} x_{3}'  \right).
\end{align}

After some algebra, we get
\begin{align}
& g' = \frac{\beta_{g}}{2R}  \left( \frac{f^2}{f_{R}^2 R - f_{R}f - f_{RR}fR} \right) \left( \frac{x_{3}'x_{2} - x_{2}' x_{3}}{x_{2}^2} \right),  \label{deq5}  \\
& \lambda ' = \frac{\beta_{\lambda}}{2R} \left( \frac{f^2}{f_{R}^2 R - f_{R}f - f_{RR}fR} \right) \left( \frac{x_{3}'x_{2} - x_{2}' x_{3}}{x_{2}^2} \right), \label{deq6}
\end{align}
where $x_{i}' \equiv x_{i}'(x_{i}, g, \lambda)$ through the relevant evolution equation. 
The complete dynamical set of equations is now (\ref{deq1})-(\ref{deq4}) supplemented with (\ref{deq5})-(\ref{deq6}). Notice that any fixed point of (\ref{deq1})-(\ref{deq4}) automatically satisfies (\ref{deq5})-(\ref{deq6}) as well. 

One should be reminded here that the derivatives with respect to $R$, e.g $f_{R}$, can be explicitly expressed using (\ref{RG_action_2}), (\ref{fR-RG}) and (\ref{fRR-RG}). In addition, both $r$ and $m$ are implicit functions of curvature $R$, through $r \equiv r(\lambda(R), g(R))$ and $g \equiv g(\lambda(R), g(R))$. 

The RG improved dynamical system with $x_{4} = 0$ has three cosmological fixed points: An early time de Sitter, a matter, and a late time de Sitter point respectively. Of course, we expect that a radiation fixed point will appear by the time we introduce $x_{4}$. For a complete analysis and the fixed point structure and their stability one can refer to Ref \cite{Amendola_fR-DS}.

The de Sitter point $P_{1}$, the matter point $P_{5}$ and the radiation point $P_{7}$, are given in the general form $P = (x_{1}, x_{2}, x_{3}, x_{4})$ as,
\begin{align}
& P_{1} = (0, 1, -2, 0 ), \\
& P_{5} = \left( \frac{3 m_{0}}{m_{0}+1}, -\frac{4 m_{0}+1}{2 (m_{0}+1)^2}, \frac{4 m_{0}+1}{2 (m_{0}+1)} ,0 \right), \label{MatterFPCoord} \\
&P_{6} = \left (\frac{4 m_{0}}{m_{0}+1}, -\frac{2 m_{0}}{(m_{0}+1)^2}, \frac{2 m_{0}}{m_{0}+1}, \frac{-5 m_{0}^2-2 m_{0}+1}{(m_{0}+1)^2} \right) .
\end{align}

The de Sitter point $P_{1}$ is characterised by $r = -2$, and is stable as long as
\be
0<m|_{r=-2}<1. \label{cond:DS_dS_Stab}
\ee
On the other hand, the points $P_5$ and $P_6$ define a family of fixed points parametrized by $m$, all lying on the line $m = -r -1$. An acceptable matter era requires that standard GR is recovered, i.e $m \rightarrow 0$ ($f_{RR} \simeq 0$), yielding $P_{5} = (0, -1/2, 1/2)$, and therefore $r = -1$.  For $m \simeq 0$, and in the presence of radiation, a radiation fixed point will also exist in the vicinity of $P_{5}$. In particular, the existence of a saddle matter era requires that at the matter point,
\be
 \left. m \right|_{r=-2}  \simeq + 0, \; \; \left. \frac{dm(r)}{dr}\right|_{r=-2} >-1. \label{cond:DS_Matter_Stab}
\ee

\begin{center}
\begin{figure}
 \includegraphics[scale=0.5]{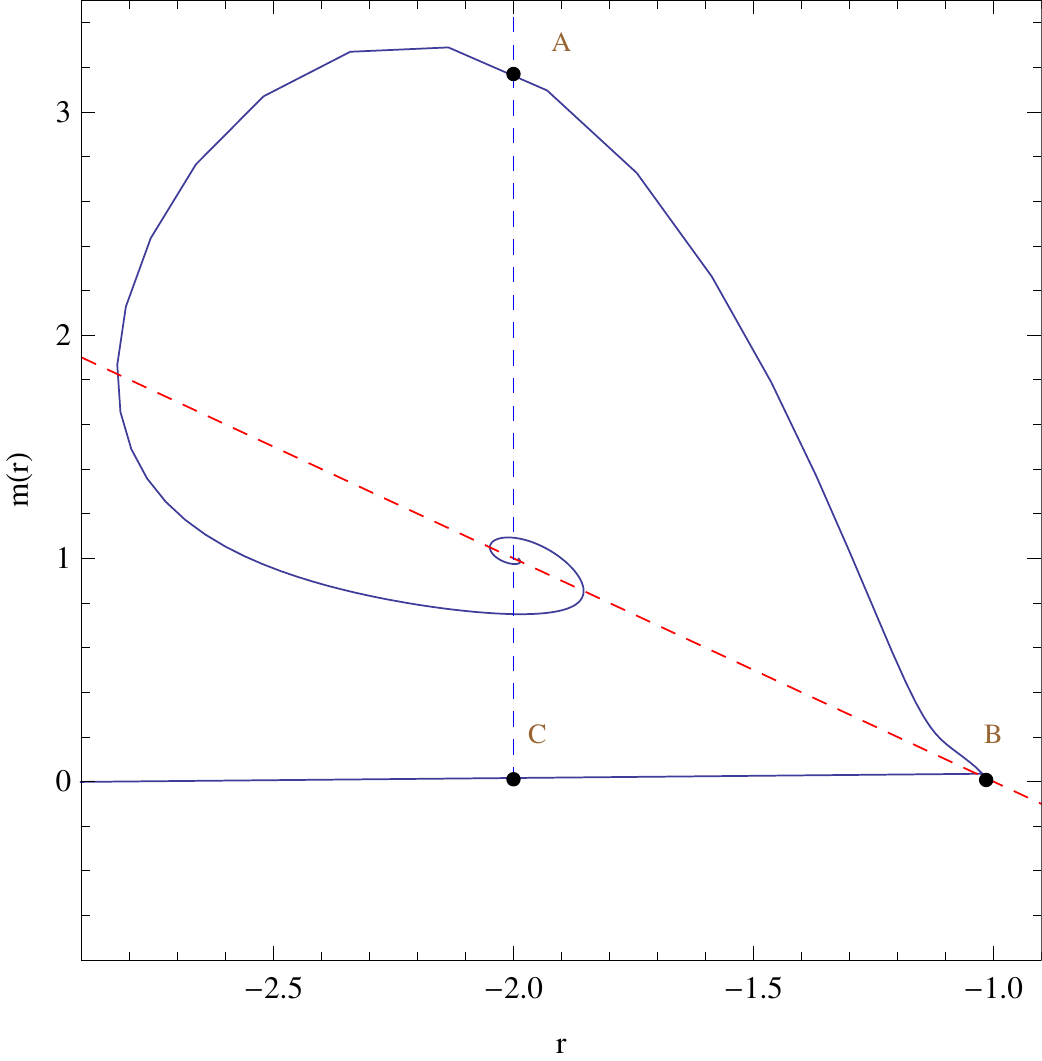}
 \caption{ \label{plot:PmrCompl} The $m-r$ plane for $\rho = 1$ and the set of initial conditions (\ref{cond:RG_IC}), with $m(r)$ and $r$ given by relations (\ref{m(r)}) and (\ref{r}) respectively. Point A corresponds to the unstable UV de Sitter point, point B to the saddle matter point, while C to the stable IR de Sitter respectively, as described in section \ref{dynamical-system}. The dashed lines correspond to $r = -2$ and $m = -r -1$ respectively.}
 \end{figure}
\end{center}

\begin{center}
\begin{figure}
 \includegraphics[scale=0.485]{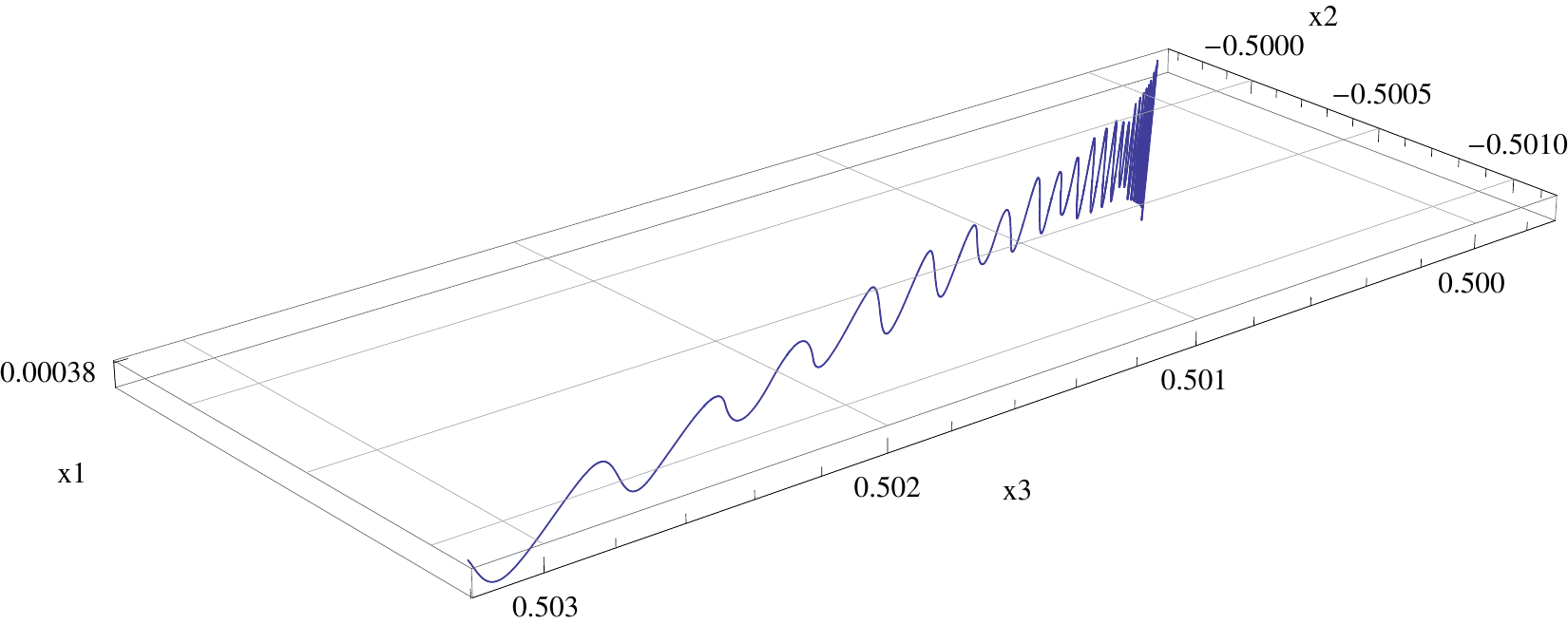} 
 \caption{\label{plot:PDSRG3D_Matter_dS} The cosmological trajectory described by the dynamical system (\ref{deq1}) - (\ref{deq4}), in the space of the coordinates $(x_1, x_2, x_3)$, leaving the matter point and evolving towards the IR de Sitter. In particular, it spirals around the unstable matter point, and then evolves towards the stable de Sitter in the IR. The initial conditions chosen are $(\lambda_0, g_0) = (10^{-2}, 10^{-5})$, and $(x_{10}, x_{20}, x_{30}) = (x_{10({\rm m})} + 10^{-5} , x_{20({\rm m})} - 10^{-6}, x_{30({\rm m})} + 10^{-5})$, with $(x_{10({\rm m})} , x_{20({\rm m})}, x_{30({\rm m})})$ denoting the coordinates of the matter fixed point given in (\ref{MatterFPCoord}), and $m_{0}$ is evaluated as $m_{0}(\lambda_0, g_0)$ using (\ref{m(r)}) and (\ref{fR-RG})-(\ref{fRR-RG}). Above initial conditions give $r_{0} \simeq -1.02$, $m_0 \simeq 1.24\times 10^{-4}$. We also assumed that $x_{4} = 0$. The amplitude of the oscillation along the $x_1$ axis is of the order $10^{-5}$.}
 \end{figure}
\end{center}

The shape of the curve $m=m(r)$ on the $m-r$ plane can provide us with sufficient information regarding the asymptotic behavior of the particular $f(R)$ model. 
In our case, we can work out the $m=m(r)$ curve by integrating the system of beta functions, and then evaluating both $r = r(\lambda, g)$ and $g=g(\lambda, g)$. By choosing a typical RG trajectory for $\rho = 1$ (i.e $k^2 = R$), and initial conditions for the system of beta functions those of (\ref{cond:RG_IC}), we get the $m-r$ curve shown in Fig.\  \ref{plot:PmrCompl}.
We see that cosmological evolution begins from an unstable ($r > 1$) early time de Sitter point, and then evolves towards the (radiation) matter point at $(r, m) \simeq (-1,0)$. It then leaves the matter point and evolves towards a stable IR de Sitter point at $r = -2$. Notice that the matter point is approached from positive values of $m$ as condition (\ref{cond:DS_Matter_Stab}) requires. 
For illustrative purposes, Fig.\  \ref{plot:PDSRG3D_Matter_dS} shows the cosmological evolution from the matter to the IR de Sitter point in the coordinate space, while Fig.\  \ref{plot:PInflMatt} shows the evolution of the effective index and slow roll parameter $w_{\rm eff}$ and $\epsilon_{V}$ respectively, from the UV de Sitter point to the matter one.   

\begin{center}
\begin{figure}
 \includegraphics[scale=0.6]{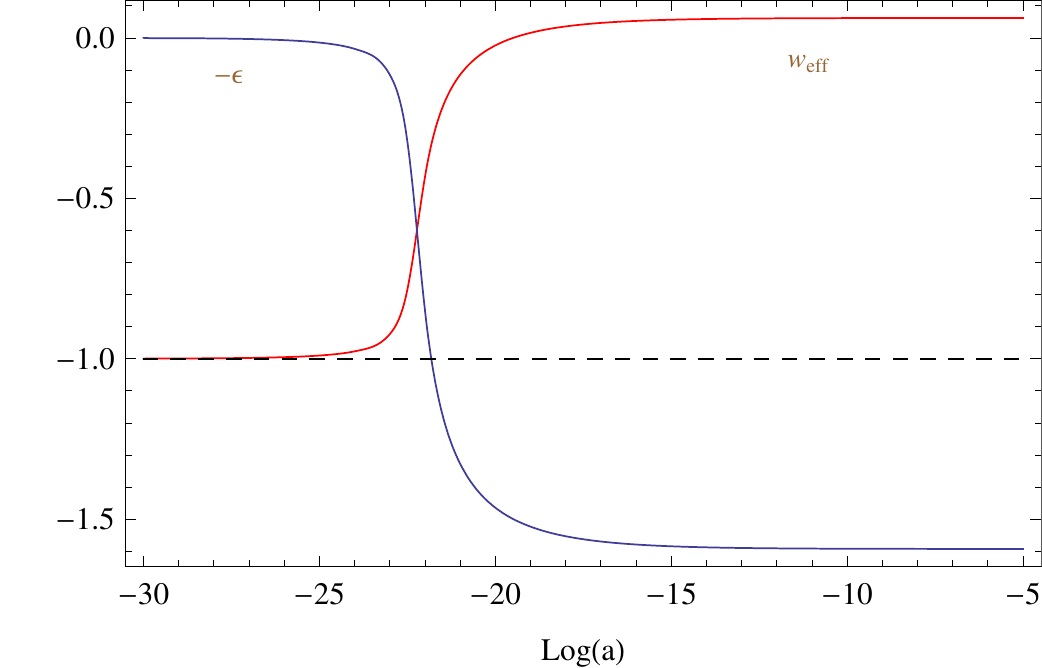} 
 \caption{The effective index $w_{\rm eff}$ and the slow roll parameter $\epsilon$ (relations (\ref{weff}) and (\ref{epsilonV}) respectively) from the UV de Sitter to matter domination for initial conditions: $(x_{10},x_{20},x_{30},\lambda_0,g_0) = (10^{-2},  -1 - 10^{-3}, 2 - 10^{-5}, 0.26, 0.02 )$ and $\ln(a_0) = -30$, and $\epsilon$ re-expressed as $\epsilon = 2 - x_{3}$.}
 \end{figure}
\end{center}
As it turns out, under a suitable choice of initial conditions for $g$, $\lambda$ and $\rho$, it is possible to get a cosmology where the UV regime is correctly connected with the IR one. The question that arises is if there are any bounds on the parameter $\rho$ in this direction. In fact, for $\rho \gtrsim 1.1$ the behavior of the evolution on the $m-r$ plane starts becoming unstable, and evolution does not reach the late time de Sitter point, after leaving the matter era. What is more, as $\rho$ increases the matter era happens to be approached from negative values of $m$, which as explained before is forbidden. Furthermore, as was also explained in the previous section, the positivity of $m_{\rm eff}^2$ in the IR regime (stability of IR de Sitter point) as well as in the GFP regime puts the extra restriction $ 0.9 \lesssim  \rho < 2$. 

Therefore, we conclude that the viability of both the classical regime and late time cosmology restricts $\rho$ to lie in the range
\be
0.9 \lesssim \rho \lesssim 1.1. \label{Rho-condition}
\ee 

\subsection{Inflationary dynamics} \label{sec:Inflation}

We showed that our particular $f(R)$ model exhibits an unstable UV de Sitter point, which can be dynamically connected with the radiation/matter era in a viable way. We would like to understand if the UV de Sitter point, describing a primordial inflationary era, could be observationally viable i.e if the scalar and gravitational fluctuations amplitudes as well as the number of e-foldings are those that are required according to observations. Recall that the only free parameter in our model is the dimensionless parameter $\rho$.

Below, we will evaluate all inflationary quantities in the Einstein frame, ignoring the non-minimal coupling between matter and the scalar field, since inflation is a (almost) vacuum dominated period.

Let us first revise some standard notions of scalar field inflation. To start with, the slow roll parameters ensure that the scalar field (inflaton) has a small kinetic energy during inflation, compared to the potential energy, so that the latter dominates. The two slow roll parameters are defined as
\begin{align}
& \epsilon \equiv \frac{\dot{H}}{H^2} = - \frac{d \ln H}{dN}, \\
& \eta \equiv \frac{\ddot{\Phi}}{H \dot{\Phi}} = \epsilon - \frac{1}{2 \epsilon} \frac{d \epsilon}{dN},
\end{align}
with the overdot denoting differentiation with respect to cosmic time.
For a scalar field action with a kinetic term and a potential, they  can be alternatively (and equivalently to first order in $\epsilon$, $\eta$) defined as
\begin{align}
&\epsilon_{V} = \frac{m_{p}^2}{16 \pi}\left(\frac{U_{\Phi} }{U}\right)^2 \label{epsilonV}, \\
& \eta_{V} = \frac{m_{p}^2}{8 \pi} \frac{U_{\Phi \Phi}}{U},
\end{align}
with the subscript $\Phi$ denoting differentiation with respect to the Einstein frame scalar field $\Phi$ respectively. 

Inflation occurs as long as the slow roll condition is satisfied, i.e
\be
\epsilon_{V} \ll 1, \eta_{V} \ll 1,
\ee
and ends when $\epsilon_{V}, \eta_{V} \sim O(1)$. Smallness of $\epsilon_{V}$ ensures that the spacetime during inflation remains sufficiently close to de Sitter, while smallness of $\eta_{V}$ ensures that variation of $\epsilon_{V}$ per e-fold is sufficiently small. 

The number of e-folds is given by
\be
N \equiv \ln\frac{a_{\rm f}}{a_{\rm i}} \approx \int_{\Phi_{\rm i}}^{\Phi_{\rm f}} \frac{U}{U_{\Phi}} d \Phi, \label{e-folds}
\ee
with $a_{\rm i}$, $a_{\rm f}$ the scale factor at the start and end of inflation respectively, and the slow roll approximation used in the last approximation. Above integral can be of course evaluated in terms of the couplings and curvature $R$ through,
\be
d\Phi = \left(\frac{\partial \Phi}{\partial g}\frac{dg}{dR} +  \frac{\partial \Phi}{\partial \lambda} \frac{d \lambda}{dR} +  \frac{\partial \Phi}{\partial R} \right) dR,
\ee
and the integral (\ref{e-folds}) can be calculated between two points $R_{\rm i}$ and $R_{\rm f}$ along the RG trajectory.
Notice that in the vicinity of a de Sitter point the number of e-folds diverges since there $U_{\Phi} \to 0$.

\begin{figure}
\centering
\begin{tabular}{cc}
\includegraphics[scale=0.4]{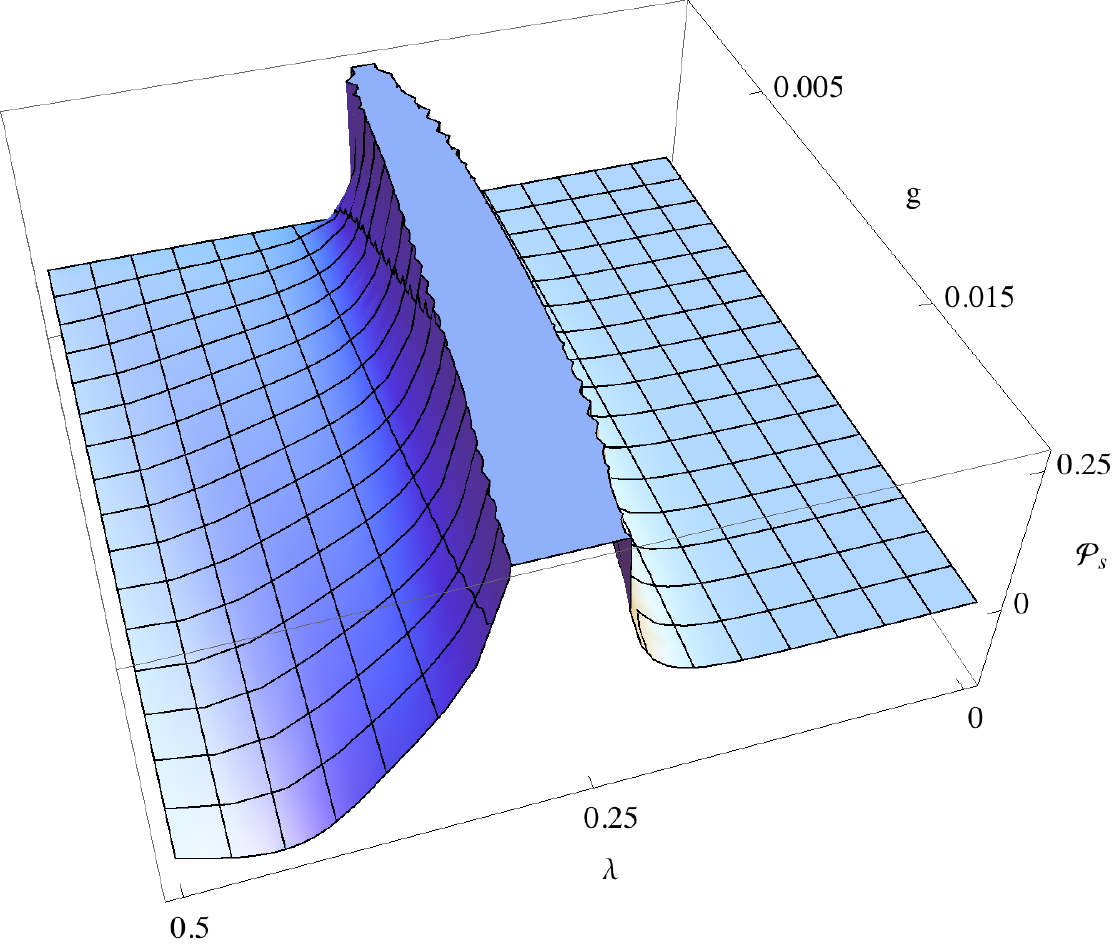} \label{PSpectra3D} & \includegraphics[scale=0.4]{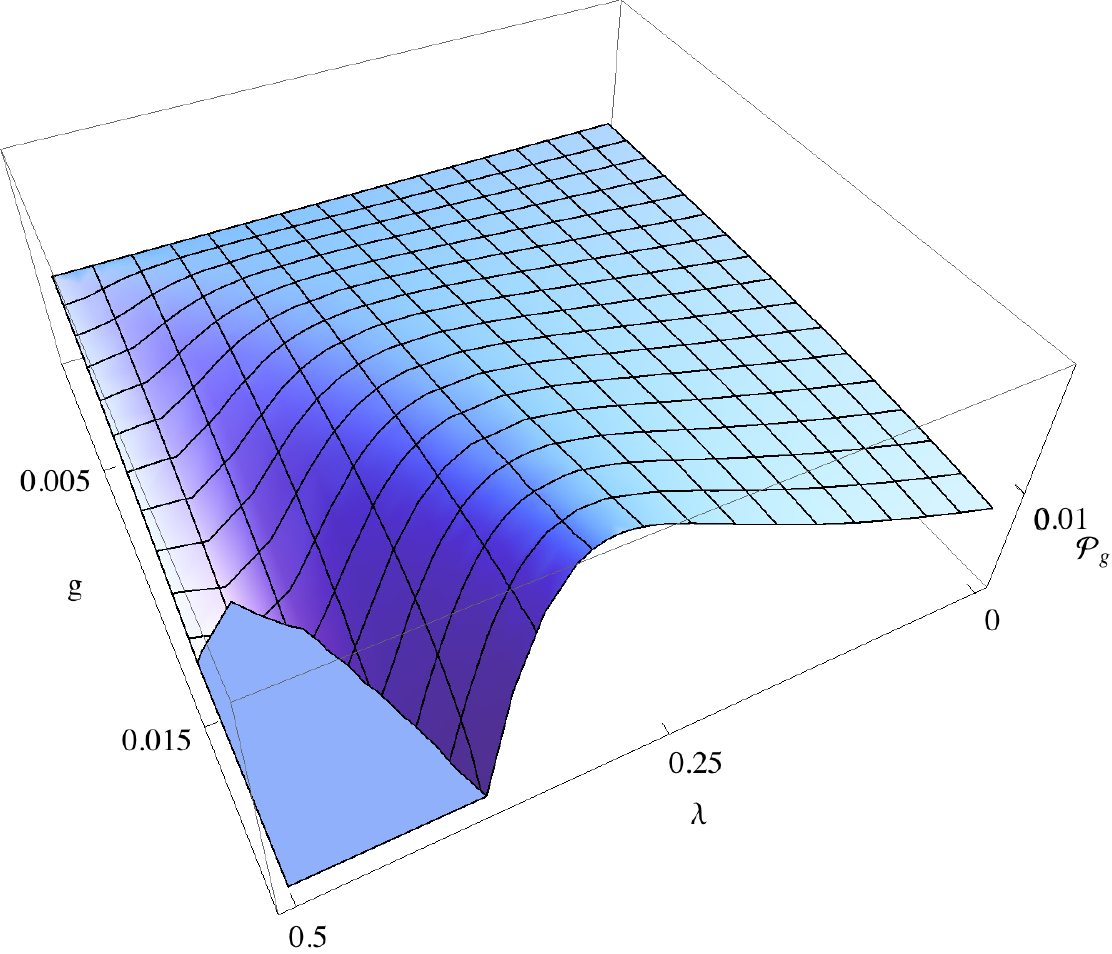} \\
\includegraphics[scale=0.4]{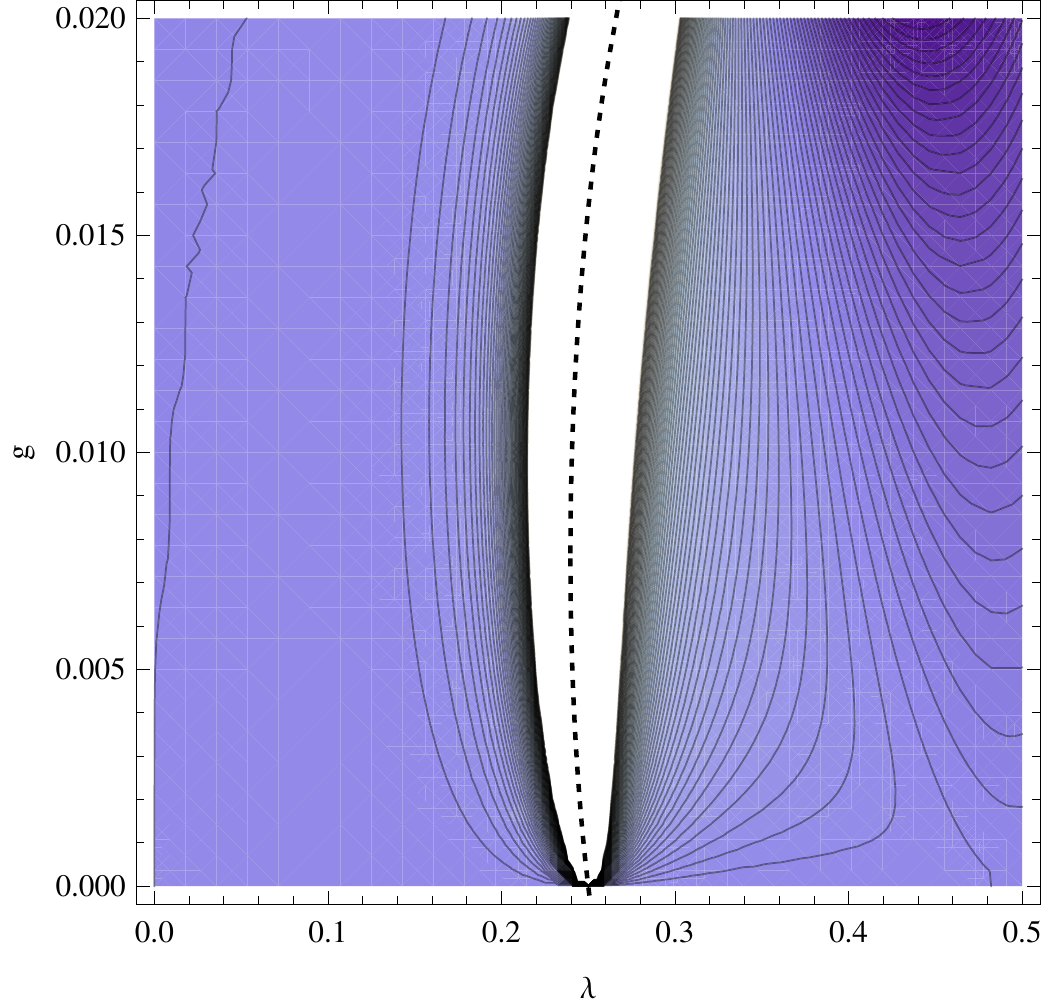} \label{CPSpectra} & \includegraphics[scale=0.4]{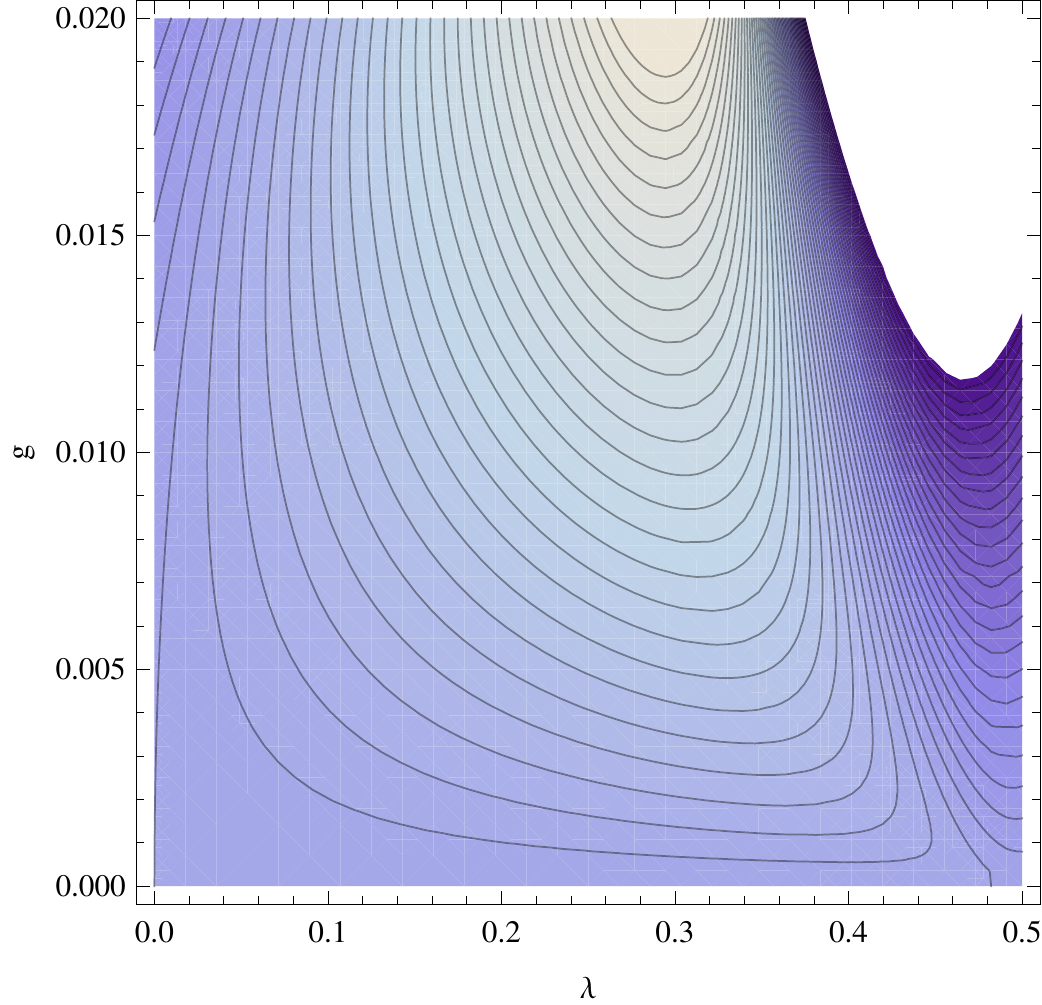}
\end{tabular}
\caption{\label{plot:PInflMatt} Upper row: The scalar (left) and gravitational (right) fluctuation power spectrum, as given by relations (\ref{rel:Scalar_Ampl_Expl}) and (\ref{rel:Grav_Ampl_Expl}) respectively, as a function of the couplings $\lambda, g$, and setting $\rho = 1$. The scalar power spectrum peaks along the de Sitter line, as on a de Sitter point it is $ {\cal{P}}_{s} \to \infty$. Lower row: The corresponding contour plots of the scalar (left) and gravitational (right) spectrum of upper row. In the scalar power spectrum the dotted line corresponds to the de Sitter line, along which the power spectrum diverges. Higher values correspond to lighter shaded areas. }
\end{figure}

Fluctuations of the scalar field during inflation, generate scalar and gravitational perturbations, whose power spectra in the slow roll approximation are given by (see e.g. \cite{Lyth:2009zz})
\begin{align}
&{\cal{P}}_{s} = \frac{128\pi}{3} \left. \frac{U^{3}}{m_{p}^6U_{\Phi}^{2}}\right|_{k = aH}, \label{def:Scalar_Ampl}\\
&{\cal{P}}_{g} = \frac{128}{3} \left. \frac{U}{m_{p}^4}\right|_{k = aH}, \label{def:Grav_Ampl}
\end{align}
assuming evaluation at the horizon crossing of the relevant mode. The scalar power spectrum becomes infinite when evaluated on a de Sitter point, 
reflecting the standard infra-red divergence.
This behavior can be seen in Fig.\  \ref{plot:PInflMatt}.

Notice that expressing the derivative of the potential as 
\be
\frac{\partial U}{\partial \Phi} = \frac{\partial U}{\partial R} \frac{\partial R}{\partial \Phi} =  \sqrt{ \frac{16 \pi G_{0}}{3} } \frac{f_{R}}{f_{RR}}  \frac{\partial U}{\partial R}, \label{rel:Potential_chain-rule}
\ee  
and using a similar expression for the second derivative, the slow roll parameters in the Einstein frame can be calculated to be
\begin{align}
& \epsilon_{V}(R) =  \frac{1}{3} \left( \frac{2f - R f_{R}}{f - R f_{R}} \right)^2 , \\
& \eta_{V}(R) = \frac{2}{3} \frac{f_{R}^2 + f_{RR}f_{R}R - 4f_{RR}f  }{f_{RR}(Rf_{R} -  f)}.
\end{align}
These relations can also be viewed as a function of the scalar $\Phi=\Phi(R)$, through relation (\ref{Scalar_Einstein}), as well as functions of $g, \lambda, \rho$ through relations (\ref{fR-RG}) and (\ref{fRR-RG}).  

The equation $\epsilon_{V}(g, \lambda) = 1$ defines a curve in the $g-\lambda$ plane (``slow roll line"), whose intersection with the RG phase curve corresponds to the end of inflation, and is associated with the corresponding de Sitter line for a given $\rho$. 
The slow roll line for $\rho = 1$ can be seen in Fig.\ \ref{plot:PPhSpdSMassZero}.

In general, decreasing $\rho$, the slow roll lines shift away from the UV RG fixed point along the RG evolution, and vice versa as $\rho \rightarrow \infty$. The opposite is true for the de Sitter lines, which means that an increasing $\rho$ increases (decreases) the scale where inflation starts (ends), and the opposite is true for decreasing $\rho$. It is interesting to note that for $\rho \rightarrow \infty$, the low energy de Sitter point lies before the point where $\epsilon_V = 1$. The limiting slow roll lines for $\rho \rightarrow 0, \infty$ are shown in Fig.\ \ref{plot:PLimdSEpsilonV}.

Let us now move to the power spectra, given in (\ref{def:Scalar_Ampl}) and (\ref{def:Grav_Ampl}). In order to match the scalar fluctuation amplitude according to the CMB observations \cite{Komatsu_WMAP7}, we need ${\cal{P}}_{s}  \simeq 2\times10^{-9}$, and ${\cal{P}}_{g}  \lesssim 0.2 {\cal{P}}_{s}$. The precise value of the amplitudes depends on a set of values for ($g, \lambda, \rho$) evaluated at the particular scale of interest. It will be useful first to give the explicit expressions of the spectra in terms of $g, \lambda$ and $\rho$, for the beta functions (\ref{betafunc1}, \ref{betafunc2}). We find 

\begin{align}
&{\cal{P}}_{s} = \frac{128}{3 \rho} \frac{ A (g, \lambda, \rho) B (g, \lambda, \rho)^3}{ C(g, \lambda, \rho)^2 D(g, \lambda, \rho)^2}, \label{rel:Scalar_Ampl_Expl}\\
& {\cal{P}}_{g}  = \frac{128}{ \rho} \frac{ A(g, \lambda, \rho) B(g, \lambda, \rho)}{ C(g, \lambda, \rho)^2} \label{rel:Grav_Ampl_Expl} ,
\end{align}   
with the additional definitions
\begin{align}
&A(g, \lambda, \rho) \equiv g \left( 4 g-(1-2 \lambda )^2 \right),\nonumber  \\
&B(g, \lambda, \rho) \equiv  96 g^2 \rho +g \left(\left(-24 \lambda ^2+4 \lambda +6\right) \rho -6\right) \nonumber \\ 
& \hspace{1.83cm} -(1-2 \lambda )^2 \lambda  \rho, \nonumber \\
&  C(g, \lambda, \rho) \equiv -192 g^2 \rho +4 g \left(3 \left(4 \lambda ^2-1\right) \rho +2\right) \nonumber \\
& \hspace{1.85cm} +(1-2 \lambda )^2 \nonumber \\
& D(g, \lambda, \rho) \equiv   -192 g^2 \rho +4 g \left(\left(12 \lambda ^2-4 \lambda -3\right) \rho +4\right) \nonumber \\ 
 & \hspace{1.85cm} +(1-2 \lambda )^2 (4 \lambda  \rho -1).
\end{align}
We arrived at relations (\ref{rel:Scalar_Ampl_Expl})-(\ref{rel:Grav_Ampl_Expl}), using relations (\ref{rel:Potential_chain-rule}) and (\ref{fR-RG})-(\ref{fRR-RG}) to re-express the spectra appropriately.  Analgous (but more complicated) expressions can be derived for beta functions with other gauges and cut-off functions. 

We have seen in the previous sections that stability requirements of the classical regime (GFP regime) as well as of the late time cosmology require that $0.9 \lesssim \rho \lesssim 1.1$. Therefore, the first thing to investigate is inflation can be observationally viable for $\rho$ in this range. 

So, let us proceed by studying the case of $\rho =1$. In this case, we also know the values of the couplings at which inflation starts and ends , $P_{\rm start} \equiv (g_{\rm start},\lambda_{\rm start}) \simeq (0.02, 0.27)$ and $P_{\rm end} \equiv (g_{\rm end}, \lambda_{\rm end}) \simeq (0.02, 0.22)$, with $P_{\rm start}$ corresponding to the UV de Sitter point, and $P_{\rm end}$ to the point where $\epsilon_{V} =1$ (see also Fig.\ \ref{plot:PPhSpdSMassZero}). 
For the connection with observations one is in principle interested at the value of the power spectra about $60$ e-foldings before the end of inflation. Now, for $\rho =1$, and as can also be seen in Fig.\ \ref{PSpectra3D}, between $P_{\rm start}$ and $P_{\rm end}$ both power spectra are smooth, decreasing functions of $g$ and $\lambda$, acquiring their lowest value at $P_{\rm end}$,
\be
{\cal{P}}_{s} \simeq 0.067, \; \; {\cal{P}}_{g} \simeq 0.052. \label{PowerSpectra_End}
\ee

One sees that the (lowest) values of the power spectra (\ref{PowerSpectra_End}), are too large to agree with observations, yielding a non-viable inflationary period for $\rho = 1$. It is not difficult to check that this behavior is true for all values of $\rho$ between $0.9 \lesssim \rho \lesssim 1.1$. Therefore, a viable late time cosmology cannot be combined with a viable primordial inflation.  

Having seen that an observationally viable inflationary era is not in agreement with a viable late time cosmology, which requires $\rho \sim 1$, we ask the following question: could inflation be viable on its own for some parameter $\rho$, away from $\rho \sim 1$? Let us try to understand this by checking the behavior of the power spectra (\ref{rel:Scalar_Ampl_Expl}) and (\ref{rel:Grav_Ampl_Expl}) for the extreme cases of $\rho \rightarrow 0$ and $\rho \rightarrow \infty$ respectively. Assuming a (finite) value of  $g$ and $\lambda$ we find that 
\be
\lim_{\rho \to 0} {\cal{P}}_{s},  {\cal{P}}_{g} = \infty,
\ee
which is obviously unacceptable. 

On the other extreme, i.e when $\rho \to \infty$, the power spectra go to zero,
\be
\lim_{\rho \rightarrow \infty} {\cal{P}}_{s},  {\cal{P}}_{g}  = 0,
\ee
which is potentially viable. For the scalar to tensor ratio we find that
\begin{align}
\left.\frac{ {\cal{P}}_{g} }{ {\cal{P}}_{s} }  \right|_{\rho \to \infty} = \frac{48 \left(48 g^2+g \left(-12 \lambda ^2+4\lambda +3\right)-(1-2 \lambda )^2 \lambda \right)^2}{  \left(96 g^2+g \left(-24 \lambda ^2+4 \lambda +6\right)-(1-2 \lambda )^2 \lambda \right)^2}.
\end{align}  
Remembering that when $\rho \gg 1$, the end of inflation, which is described on the phase space of $g-\lambda$ by the slow roll line, is shifted towards smaller values of the couplings, as can also be seen in Fig.\ \ref{plot:PLimdSEpsilonV}. Therefore, we can get an estimate of above ratio by assuming that the fluctuations are produced at a point in the linear regime of the RG evolution, where $g \sim \lambda \ll 1$, yielding ${\cal{P}}_{g} / {\cal{P}}_{s}  \sim O(1)$, which is observationally unacceptable.  

Before concluding this section, let us comment on another possibility of understanding inflation in this scenario, that is modeling it as $R^2$ inflation \cite{Star-R^2} using the $f(R)$ model found in (\ref{f(R)_linear_correct}) at large $R$:
\be
f(R) \simeq  \frac{\tilde\kappa^2}{G_0}\left(R - 2 \Lambda_{0} \right) +  6 ( 2 - \rho) \rho  R^2. \label{f(R)_linear_correct-repeat}
\ee
Matching to the perturbation amplitude,  $R^2$
 inflation can account for the observations if 
the coefficient of the $R^2$ term is of order $10^{11}$ \cite{StarobinskyInflation1}.
Hence we see in approximate way how tuning $\rho$ to very large values suppresses the perturbations.  
However, this results in an unacceptable classical limit as well as a non-viable late time cosmology for the reasons explained in previous sections.

To conclude this section, it turns out that primordial inflation in this scenario cannot agree with observations unless $\rho$ is very large, in which case the mass of the scalaron diverges and becomes tachyonic in the subsequent evolution. Hence the observed fluctuations must be generated at a later period of inflation, which requires that 
more degrees of freedom should be introduced in the action, like for example a scalar field. Scalar field inflation in the asymptotic safety scenario, and with scale identification in the equations of motion,  has been considered in \cite{Contillo:2011ag}.   
A more exotic possibility is that the extra degrees of freedom produce a fixed point with a very small fluctuations.  One notes that for small $g$, the tensor power spectrum becomes
\begin{equation}
{\cal{P}}_{g} \simeq 128{g\lambda}
\end{equation}
which is suggestive that a fixed point with small $g\lambda$ could be viable. Note the appearance of the product $g\lambda \sim G\Lambda$, which is the expected scale of tensor fluctuations in Einstein-Hilbert gravity in a de Sitter phase with cosmological constant $\Lambda$.


\section{Discussion and conclusions}
We studied the cosmology of an $f(R)$ model generated by the RG improvement of the Einstein--Hilbert action.
The transition to $f(R)$ gravity was achieved by identifying the renormalisation group scale to be proportional to scalar curvature, 
\be
k^2 = \rho R,
\ee
in the non-perturbative beta-functions calculated from the exact renormalisation group equation. 

We found that the resulting $f(R)$ model has some remarkable properties. 
Firstly, it maintains the correct sign for the graviton and scalaron kinetic terms.
Very close to a non-trivial RG fixed point it behaves like $R^2$ gravity, which is scale invariant,  while it reduces to GR in the vicinity of the Gaussian fixed point. 
At solar and galactic scales, the scalaron's mass is of the order of Planck mass, preventing observable departures from GR at these scales. 
On the other hand, in the vicinity of the UV RG fixed point, the scalaron mass vanishes, reflecting the scale invariance of the action in that regime. 

The cosmological solutions of the $f(R)$ model are also interesting.  It naturally exhibits an unstable UV de Sitter point which evolves to a stable one in the IR. The effective cosmological constants are exponentially separated when Newton's $G$ and the cosmological constant are matched to their observed values. What is more, there are an infinite set of de Sitter points as the UV RG fixed point is approached ($R \to \infty$). However, classical cosmological evolution starts from the outermost de Sitter point, and therefore the UV RG fixed point is hidden behind it, and cannot be accessed. The Big Bang singularity is avoided, since the de Sitter point is reached at infinite time in the past, i.e as $t \to - \infty$. 
The model therefore satisfies the requirements of a successful $f(R)$ model itemized in Ref.\ \cite{Odin_Noji-Unification1}.

Introducing matter content to the cosmology, we found that the UV de Sitter point can be connected to the IR de Sitter era through a radiation/matter era, with a stable scalaron, provided  
\be
0.9 \lesssim \rho \lesssim 1.1.
\ee 
Unfortunately, the fluctuations generated during inflation at the outer UV de Sitter point are too large to account for the observations (Section \ref{sec:Inflation}). 

Therefore observable inflation requires extra degrees of freedom in the action, for example a scalar field driving inflation at a lower scale. A more remote possibility would be that the extra degrees of freedom move the fixed point to a smaller value of $g\lambda$, which could suppress the fluctuations. 

To make the comparison with previous cut-off identifications in the literature, performed at the level of the equations of motion, our constraint for the parameter $\rho$, i.e $\rho \sim 1$, is broadly consistent with scale identifications made in the equations of motion, rather than the action as here. In particular, in Ref. \cite{Reuter-Sauer_RG-Cosmo}, it was numerically found that for the identification $k^2 \sim c H^2$, the constant $c$ should be of order one, which is consistent with $\rho \sim 1$ in our identification.

{
This model can be improved by extending the analysis performed in this paper to higher truncations, i.e by including higher order curvature terms in the action. It is interesting to ask what features are generic.  The existence of a UV fixed point seems to be a universal feature of all truncations found so far, so we expect the Einstein frame potential of the scalaron to tend to a constant at large values of the field.  However, we do not expect the presence of an infinite number of de Sitter points to be generic, as it arose from the complex eigenvalues of the fixed point, which are not present for the general four-derivative truncation \cite{Benedetti:2009rx}.
We should also include matter fields in the renormalisation group equations.  
With these modifications it might turn out that there is a model for which 
both early and late time cosmology agrees with observations. 
}

\textit{Note added.} While this work was being finalised, Ref. \cite{Bonanno:2012jy} appeared, which also makes the scale identification in the action, and finds an infinite number of de Sitter solutions near the UV fixed point.

 \acknowledgements
We would like to thank Daniel Litim and Christoph Rahmede for reading the manuscript and providing us with constructive feedback. We acknowledge support from the Science and Technology Facilities Council [grant number ST/J000477/1].



\end{document}